\documentclass[a4paper]{article}

\usepackage[pages=all, color=black, position={current page.south}, placement=bottom, scale=1, opacity=1, vshift=5mm]{background}
\SetBgContents{
	\tt  
}      

\usepackage[margin=1in]{geometry} 

\usepackage{amsmath}
\usepackage{amsthm}
\usepackage{amssymb}

\usepackage[utf8]{inputenc}

\RequirePackage[colorlinks,citecolor=blue,urlcolor=blue]{hyperref}
\hypersetup{
	unicode,
	pdfauthor={Federica Zoe Ricci},
	pdftitle={Bayesian temporal biclustering with applications to multi-subject neuroscience studies},
	pdfsubject={A simple article template},
	pdfkeywords={article, template, simple},
	pdfproducer={LaTeX},
	pdfcreator={pdflatex}
}

\theoremstyle{plain}

\theoremstyle{definition}

\usepackage{graphicx, color}
\graphicspath{{fig/}}

\usepackage{algorithm, algpseudocode} 
\usepackage{mathrsfs} 

\usepackage{lipsum}

\RequirePackage{amsthm,amsmath,amsfonts,amssymb, soul}
\RequirePackage[authoryear, round]{natbib}

\RequirePackage[colorlinks,citecolor=blue,urlcolor=blue]{hyperref}
\RequirePackage{graphicx}

\DeclareMathOperator{\iid}{\overset{\text{iid}}{\sim}}
\DeclareMathOperator{\ind}{\overset{\text{ind}}{\sim}}

\newcommand{\CellWithForceBreak}[2][c]{
\begin{tabular}[#1]{@{}c@{}}#2\end{tabular}}

\usepackage{amsmath}        
\usepackage{bbm}            
\usepackage{amsfonts}       
\usepackage{nicefrac}       
\usepackage[many]{tcolorbox}
\newtcolorbox[]{algorithm1}{breakable, enhanced jigsaw, rounded corners, parbox=false, sharp corners, colback=white, colbacktitle=white,coltitle=black}

\usepackage{array}
\usepackage{adjustbox}
\usepackage{tikz}
\usepackage{tikzsymbols}
\usetikzlibrary{arrows, arrows.meta, automata, backgrounds, bayesnet, fit, positioning, calc}
\tikzset{
    state/.style={
           rectangle,
           rounded corners,
           draw=black, thick,
           minimum height=2em,
           inner sep=10pt,
           text centered,
           },
}

\usepackage{hyperref}
\hypersetup{
    colorlinks,
    citecolor=blue,
    filecolor=blue,
    linkcolor=black,
    urlcolor=black,
}
\usepackage{url}

\usepackage{xcolor}        
\usepackage{color}

\definecolor{lightyellow}{HTML}{ffffa7}
\definecolor{col_plate_R}{HTML}{61c1cf}
\definecolor{col_plate_Z}{HTML}{eb4702}
\definecolor{col_plate_N}{HTML}{02eb12}
\definecolor{col_plate_mu}{HTML}{b547e6}
\usepackage{bbm}

\usepackage{subcaption}

\title{Bayesian temporal biclustering with applications to multi-subject neuroscience studies}
\author{Federica Zoe Ricci$^1$\thanks{Supported by the Hasso-Plattner-Institute Research Center in Machine Learning and Data Science at UC Irvine.} \and Erik B. Sudderth $^1$\thanks{Partially supported by National Science Foundation Robust Intelligence Award No.~IIS-1816365.} \and Jaylen Lee$^1$ \and Megan A. K. Peters $^1$\thanks{Partially supported by a program of the Canadian Institute for Advanced Research Fellowship in the Brain, Mind, and Consciousness.} \and Marina Vannucci$^2$ \and Michele Guindani$^3$\thanks{Partially supported by a grant from the Karen Toffler Charitable Trust. }}

\date{
	$^1$University of California, Irvine \\ \texttt{\{fzricci, sudderth, jaylenl, megan.peters\}@uci.edu}\\%
	$^2$Rice University\\ \texttt{marina@rice.edu}\\
 	$^3$University of California, Los Angeles\\ \texttt{mguindani@g.ucla.edu}\\[2ex]%
}

\begin{document}
\maketitle
	
\begin{abstract}
We consider the problem of analyzing multivariate time series collected on multiple subjects, with the goal of identifying groups of subjects exhibiting similar trends in their recorded measurements over time as well as time-varying groups of associated measurements.
To this end, we propose a Bayesian model for temporal biclustering featuring nested partitions, where a time-invariant partition of subjects induces a time-varying partition of measurements.
Our approach allows for data-driven determination of the number of subject and measurement clusters as well as estimation of the number and location of changepoints in measurement partitions.
To efficiently perform model fitting and posterior estimation with Markov Chain Monte Carlo, we derive a blocked update of measurements' cluster-assignment sequences.
We illustrate the performance of our model in two applications to functional magnetic resonance imaging data and to an electroencephalogram dataset. 
The results indicate that the proposed model can combine information from potentially many subjects to discover a set of interpretable, dynamic patterns.
Experiments on simulated data compare the estimation performance of the proposed model against ground-truth values and other statistical methods, showing that it performs well at identifying ground-truth subject and measurement clusters even when no subject or time dependence is present.
		
\noindent\textbf{Keywords:} probabilistic clustering, nested partitions, separate exchangeability, sparse finite mixtures, fMRI, event related potentials
\end{abstract}

\hypertarget{introduction}{%
\section{Introduction}\label{sec:introduction}}

\begin{figure}[t]
    \centering
    \includegraphics[width=14cm, height=6cm]{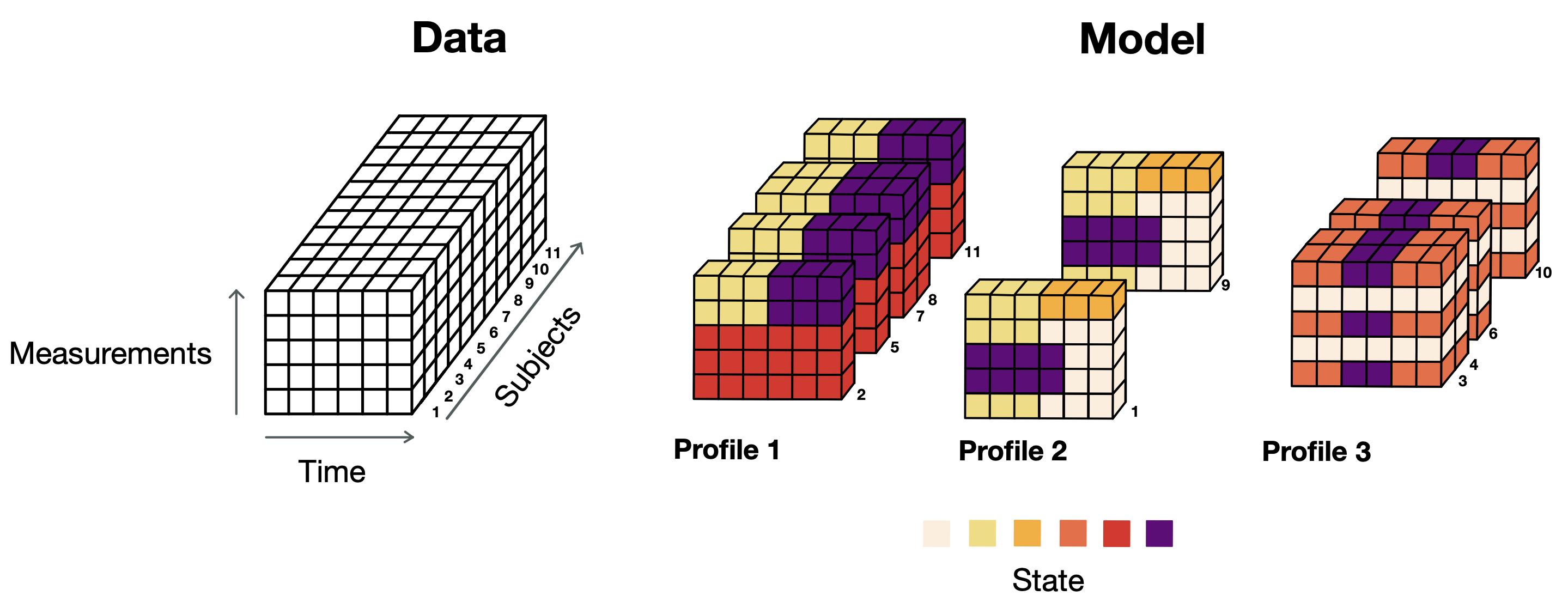}
    \caption{Illustration of proposed biclustering model for an idealized scenario with 11 subjects, 5 measurements and 6 time steps. (Left) Data displayed as a three-dimensional array. A cell represents an observation for a specific subject, measurement and time step. (Right) The proposed model clusters subjects into profiles and measurements into states. A profile is a specific configuration of state sequences over time shared by a group of subjects. The state assigned to an observation determines its likelihood parameters. In this example, there are 3 profiles and 6 states. As suggested by the figure, multiple subjects may be associated with the same profile, states can be shared across profiles, states are likely to persist over time and the probability of state changes is shared across measurements, within a profile.}
    \label{fig:data_vs_model}
\end{figure}

Many scientific studies follow multiple units over time, systematically collecting a range of measurements on each unit at set intervals. This process generates multivariate time series data for each unit. For example, in climate studies, meteorological and environmental variables are recorded at multiple locations (units), recurrently over a period of time \citep{levy_meta-analysis_2012}. Similarly, in economics, it is common to analyze the evolution of multiple financial indicators across different countries (units) \citep{lakstutiene_correlation_2008}. Here, we consider applications in neuroscience where multiple brain signals, e.g. functional magnetic resonance imaging (fMRI) data measured at several brain regions, or electroencephalogram (EEG) data measured at several electrodes, are recorded on multiple subjects (units) during the course of an experiment \citep{bowman_brain_2014, luck_introduction_2014}. In these applications, the data can be represented as a three-dimensional array, with rows corresponding to time points, columns corresponding to vectors of measurements and slices indicating different units; see Figure \ref{fig:data_vs_model} (left). Analyzing such data requires addressing the heterogeneity that stems from its dimensions.  In neuroscience applications, subgroups of subjects may display similar patterns of brain activity over time. Measurements of brain activity can reveal patterns of associations across brain regions that vary over time and between subjects.

In this paper, we introduce a \emph{temporal biclustering} framework for modeling the heterogeneity in multivariate time-series data collected over multiple units. This framework simultaneously identifies clusters of subjects (\emph{profiles}) and clusters of measurements (e.g., brain signals) over time (\emph{states}). A profile characterizes a specific sequence of states shared by a group of subjects, capturing their common temporal dynamics. States reflect probabilistic features of the data, for example the expected signal strength at a particular brain region and time point. While the actual observations may differ across subjects at each time point, subjects in a profile share the same underlying pattern of transitioning between states; see Figure~\ref{fig:data_vs_model} (right). Existing Bayesian models for clustering time-series have typically focused on temporal clustering of multivariate time series for single units. For example, \citet{fox_sticky_2011} combined hidden Markov Models and the hierarchical Dirichlet Process \citep{teh_hierarchical_2006} to analyze audio sequences recorded during a meeting and infer a single cluster per time point, indicative of the speaker at that moment. Recently,  \cite{page_dependent_2022} proposed a temporal random partition model designed for analyzing time series data with multivariate measurements. Their model associates a \emph{sequence} of cluster assignments to \emph{each} measurement. A multi-unit model was proposed by \cite{ren_biclustering_2020} to cluster individuals based on their diastolic blood pressure  collected over time. In addition to being limited to a single measurement, their approach directly models the number of changepoints, necessitating the use of a computationally costly reversible-jump Markov Chain Monte Carlo (MCMC) algorithm for posterior inference.

Our focus is on multivariate time-series data collected from multiple units, with the aim of clustering patterns across both subjects and time. Our approach achieves bi-clustering through two key assumptions. First, we assume \emph{nested} partitions, whereby  a time-varying partition of measurements is associated to a time-invariant partition of subjects. Secondly, we employ an assumption of \emph{separate exchangeability} \citep{lin_separate_2021}, which states that for subjects within the same profiles, observations of measurements at the same time points are considered exchangeable.  This assumption is reasonable for measurements collected at different locations in the brain.
Consequently, at each time point, the multiple measurements within each profile share the same state, allowing for borrowing of strength both within subjects' profiles and across time points. Similar to the approach in \citet{page_dependent_2022}, we account for temporal dependence by encouraging measurements to persist in the same state over consecutive time steps while learning the number and position of changepoints from the data. For static bi-clustering, \cite{lee_nonparametric_2013} proposed the use of nested partitions. In their framework, a partition of samples is nested within a partition of proteins, and two proteins are assigned to the same cluster whenever they induce the same partition of samples. This idea was later formalized and connected to the principle of separate exchangeability by \cite{lin_separate_2021}, providing a general framework for this modeling approach. Building upon the plaid model of \cite{lazzeroni_plaid_2002}, \cite{murua_biclustering_2022} proposed a method for estimating overlapping biclusters in gene expression and histone modification data. In a more recent development, \cite{yan_bayesian_2022} proposed Bayesian biclustering models specifically designed for categorical data, with applications to genetic datasets. One of the few   methods for biclustering of three dimensional arrays was developed by \cite{mankad_biclustering_2014} as a  non-Bayesian extension of the plaid model by \cite{lazzeroni_plaid_2002}.  
When applied to time series data, their model can find potentially overlapping biclusters with time-varying effects, similar to a bicluster-specific functional ANOVA.  However, this flexibility can lead to hard-to-interpret results and impact the optimization of the model parameters, as we show in our simulation study.

 Our approach employs a robust location-scale t-distribution as the likelihood model, which accommodates heterogeneity between observations assigned to the same state while preserving interpretability and attenuating the influence of outliers.  To capture the temporal evolution of states, we adapt the temporal random partition model (tRPM) proposed by \cite{page_dependent_2022} to the multivariate measurement and multi-subject case. The approach allows for temporal persistence of measurements in the same state, promoting stability and continuity over time. Moreover, it enables flexible resampling of states from a profile-specific categorical distribution, accommodating potential changes and transitions in the underlying patterns. We leverage sparse finite mixture priors \citep{malsiner-walli_model-based_2016} to  determine the number of subject and measurement clusters --- that is, profiles and states --- in a data-driven manner. The number and location of changepoints in measurement partitions are also automatically estimated. Our framework yields interpretable results 
 summarizing information from multivariate multi-subject time-series data into a smaller set of dynamic patterns that capture their underlying structure and temporal evolution. 
 
 We also develop a novel and efficient algorithm for posterior inference. Unlike previous approaches using marginal updates \citep{page_dependent_2022}, which can lead to slow and ineffective posterior exploration, we derive a variant of the forward-backward algorithm \citep{rabiner_tutorial_1989} for blocked variable updates.  This allows for updating the entire sequence of state assignments altogether, conditioning only on the previous state assignment,  enabling the MCMC sampler to efficiently transition between state sequences over consecutive iterations.

Although our temporal biclustering framework is widely applicable to studies where multiple subjects undergo
the same experiment with multiple measurements recorded at the same time intervals, in this paper we particularly focus on its application to data from fMRI and EEG experiments. For multi-subject fMRI data, common approaches include two-stage analyses. Although approaches may vary, in these methods, estimates of model parameters are typically first summarized at the individual level (e.g. by fitting a generalized linear model (GLM) separately to each subject) and then plugged into a model for the population level analysis  \citep[e.g. performing analysis of variance on the aggregated GLM results, see][]{bowman_fmri_2008}. Few multi-subject approaches have been developed in the last decade that do not rely on a two-stage approach but analyze multiple subjects jointly. For example, \cite{zhang_spatiotemporal_2016} propose a Bayesian model for the detection of sets of brain regions that activate in response to a stimulus and are possibly different across subjects. \cite{kundu_bayesian_2023} propose a Bayesian tensor response regression approach to compare brain activity between a control and an intervention group while accounting for subject-level effects. Another approach can be found in \cite{zhao_hierarchical_2023}, who propose a hierarchical Bayesian model working with partial autocorrelations rather than in the time or frequency domains, to achieve faster computational inference. In our application to fMRI data, changepoints in measurement partitions identified by our model largely coincide with the transitions between resting and activation (or task) blocks, with groups of subjects characterized by varying degrees of BOLD signal strength. 
For EEG modeling, event-related potentials (ERPs), measuring neural activity in response to specific events (e.g. motor or cognitive stimula), are typically collected on multiple subjects and often analyzed by focusing on single subjects \citep{kallionpaa_single_2019}, or averaging observations across subjects \citep{yu_bayesian_2023}. Statistical models for the analysis of multi-subject data include \cite{yu_semiparametric_2024}, who propose a semiparametric latent ANOVA model to estimate the association between ERP characteristics and subject-level covariates, and \cite{kang_scalar_2018}, who develop a scalar-on-image regression model and use soft-thresholded Gaussian processes to study the relationship between a scalar-valued outcome and observed ERP signals; these approaches do not address clustering.   
Our results on EEG data further demonstrate the ability of our model to handle heterogeneous data by identifying dynamic patterns shared by many subjects. We also compare the performance of our method across various simulation scenarios with existing methodologies.

The reminder of the article is structured as follows. In Section \ref{sec:model} we describe the proposed multi-subject multivariate temporal biclustering (MMTB) model. Section \ref{sec:posterior} explains how our model can be fit to data via MCMC, and how posterior samples from multiple MCMC chains can be combined into summaries of subject and measurement clusters. Results obtained by fitting our temporal biclustering model to two neuroscience studies, an fMRI and an EEG experiment, are illustrated in Section \ref{sec:applications}. In Section \ref{sec:simulations} our model is validated and compared to some alternative methods on simulated data with known ground-truth clusters and likelihood parameters. We conclude in Section \ref{sec:discussion} by outlining the limitations of our model and discussing future research directions.

\hypertarget{sec:model}{%
\section{Bayesian temporal biclustering}\label{sec:model}}
Let $Y_{i,r,t}$ be the signal recorded for subject $i = 1, \dots, N$, measurement (e.g. location) $r = 1, \dots, R$, and time $t = 1, \dots, T$.
Our applications will consider brain signals measured on $N$ subjects at $R$ locations in the brain (fMRI) or on the scalp (EEG) and $T$ equally-spaced time steps throughout an experiment.
We assume that $Y_{i,r,t}$ is sampled from a probability distribution $F_{\boldsymbol{\theta}_{i,r,t}}$ identified by parameters $\boldsymbol{\theta}_{i,r,t}$.  To emphasize the generality of our framework, we first define our temporal biclustering model for $\boldsymbol{\theta}_{i,r,t}$, across $i$ and $r$,
then motivate the choice of a flexible but interpretable likelihood  $F_{\boldsymbol{\theta}_{i,r,t}}$. 

\subsection{Biclustering via separate exchangeability} 
We wish to cluster signals measured at different locations and subjects over the experimental conditions (see Figure \ref{fig:data_vs_model}). For this purpose, we propose a model that enforces separate exchangeability \citep{kallenberg_exchangeability_1988, lin_separate_2021} across the different dimensions of the data via the use of Bayesian nonparametric modeling. We refer to clusters of subjects as \textit{profiles} and clusters of measurements as \textit{states}.
We assume that a subject $i$ has a single profile assignment for the entire experiment, as determined by a profile assignment variable $s_i = z \in \{ 1, \dots, Z \}$. 
The value $Z$ is an upper bound to the number of profiles $Z^*$ that are active, or present in the data. In the applications of this paper we set $Z = N$ so as to allow for any number of active subject clusters while choosing a prior that encourages $Z^* < Z$, as detailed in Sec. \ref{subsec:cluster_prior}.

We assume that subjects with the same profile share the same partition of measurements at every time step.
Figure \ref{fig:data_vs_model} (right) provides an illustration of these modeling assumptions, separating slices of data according to the profile of their corresponding subject, and associating to each profile distinctive sequences of states for their measurements.
We assume separate exchangeability among observations collected at the same time step of subjects with the same profile.
Separate exchangeability preserves the known structure of the data by implying that observations of the same measurement are as or more highly correlated than observations of different measurements. 
Formally, we adapt the framework of separate exchangeability through nested partitions proposed by \cite{lin_separate_2021} to time series data. 
Letting $I_z = \left\{ i \in \{1, \dots, N \}: s_i = z \right\}$ be the collection of subjects with profile $z$, we assume 
\begin{equation}
    (Y_{i, r, t} : i \in I_z ; r = 1, \dots, R) \overset{\text{d}}{=} 
    (Y_{\rho^{(z)}_1(i), \rho_2(r), t} : i \in I_z ; r = 1, \dots, R) \label{eq:sep_exc}
\end{equation}
for each $t = 1, \dots, T$, where $\rho^{(z)}_1$ is any permutation of indices in $I_z$ and $\rho_2$ is any permutation of measurement indices. 
In terms of the time series data for subject $i$ at location $r$, which we denote by $\mathbf{Y}_{i, r} = (Y_{i, r, 1}, \dots, Y_{i, r, T})$, Eq.~\eqref{eq:sep_exc} implies that
\begin{equation}
    \mathbf{Y}_{i, r} \overset{\text{d}}{=} 
    \mathbf{Y}_{j, r} \quad \forall j \in \{1, \dots, N \}: s_j = s_i, \label{eq:same_distribution}
\end{equation}
or, in words, the distribution of the time series recorded at measurement $r$ for subject $i$ is the same as that recorded at the same measurement for any other subject $j$ with $s_i = s_j$.
This is achieved by defining a sequence of state assignment variables $\left(c_{r, 1}^{(z)}, \dots, c_{r, T}^{(z)}\right)$ for each profile $z$, where $c_{r, t}^{(z)} = k \in \{1, \dots, K\}$ and, similar to $Z$, $K$ is a specified upper bound to the number of states or measurement clusters. The number $K^*$ of states actually present in the data at least once is inferred from data, as explained in Sec.~\ref{subsec:cluster_prior}. 
In turn, the state of an observation identifies its sampling distribution, so that there exists a set of parameters $\{\boldsymbol{\theta}^*_1, \dots, \boldsymbol{\theta}^*_K\}$ such that $\boldsymbol{\theta}_{i,r,t} = \boldsymbol{\theta}^*_k$ whenever measurement $r$ at time $t$ is in state $k$.
These assumptions satisfy Eq. \eqref{eq:same_distribution} because, for any given measurement $r$, subjects with the same profile $z$ have the same sequence of likelihood parameters $(\boldsymbol{\theta}^*_{c_{r, 1}^{(z)}}, \dots, \boldsymbol{\theta}^*_{c_{r, T}^{(z)}})$. 
In Figure \ref{fig:data_vs_model} (right) this is illustrated by letting each cell of the data take a color corresponding to its latent sampling distribution, as determined by its state assignment. While seemingly restrictive, when paired with a robust likelihood distribution $F_{\boldsymbol{\theta}_{i,r,t}}$ that allows for considerable variability among its samples, our modeling assumptions enable us to combine information from many subjects to identify potentially interesting relationships among the measurements that evolve during the experiment, as demonstrated by our neuroscience applications.

\subsection{Dynamic evolution of measurements' clusters} \label{subsec:trpm}
We wish to model the relationships between the multiple measurements collected on each subject and how these vary over the course of an experiment.
In our framework, this is equivalent to modeling the evolution of $c^{(z)}_{r, t}$ over time.
In doing so, we wish to simultaneously account for temporal dependences by encouraging measurements to persist in the same state over consecutive time-steps, while also allowing for measurement clusters to change and for learning the number and position of changepoints from the data.
We adapt the temporal random partition model (tRPM) of \cite{page_dependent_2022}, proposed for modeling changes in the partition of multivariate time series, to our multi-subject case.
In particular, letting $\mathbf{c}^{(z)}_{t} = (c^{(z)}_{1, t}, \dots, c^{(z)}_{R, t})$ be the state assignment of all measurements at time $t$ for subjects in profile $z$, we assume that $(\mathbf{c}^{(z)}_{1}, \dots \mathbf{c}^{(z)}_{T})$ follows a tRPM.
This implies that the state assignment $c_{r,t}^{(z)} \in \{1, \dots, K\}$ of measurement $r$ at time $t$ is either equal to the state assignment $c_{r,t-1}^{(z)}$ of the same measurement at the previous time step, or it is resampled from a categorical distribution $\boldsymbol{\omega}^{(z)} = \left(\omega_1^{(z)}, \dots, \omega_K^{(z)}\right)$ over the $K$ states that is specific to profile $z$. 
For the first time step,
\begin{equation}
    c_{r,1}^{(z)} \mid \boldsymbol{\omega}^{(z)} \ind \text{Categorical}\left(\omega^{(z)}_1, \dots, \omega^{(z)}_K\right), \label{eq:c_first}
\end{equation}
and for $t = 2, \dots, T$ we assume the following  dynamics:
\begin{align}
    c_{r,t}^{(z)} \mid \omega^{(z)}, \gamma_{r,t}^{(z)} &\ind \gamma_{r,t}^{(z)} \, \delta_{c_{r,t-1}^{(z)}} + \left(1 - \gamma_{r,t}^{(z)}\right) \text{Categorical}(\omega^{(z)}_1, \dots, \omega^{(z)}_K),  \label{eq:c_dynamic}\\
    \gamma_{r,t}^{(z)} \mid a_{t}^{(z)} &\ind \text{Bernoulli}\left(a_{t}^{(z)}\right), \label{eq:gamma}\\
    a_{t}^{(z)} &\iid \text{Beta}(\alpha, \beta) \label{eq:a}.
\end{align}
In Eqs. \eqref{eq:c_dynamic} to \eqref{eq:a}, $\delta_{k}$ is the Dirac measure on the singleton set $\{k\}$, $\gamma_{r,t}^{(z)}$ is a persistence indicator variable that equals $1$ when measurement $r$ stays at time $t$ in the same state as it was at time $t-1$ and value $0$ if its state is resampled (fixing $\gamma_{r,1}^{(z)} = 0$), $a_{t}^{(z)}$ is a profile-$z$ specific state-persistence probability shared by all measurements at time $t$, and $\alpha$ and $\beta$ are hyperparameters controlling the desired degree of time stability. 

We choose to use the tRPM because, as is clear from Eqs. \eqref{eq:c_dynamic} to \eqref{eq:a}, it models the sequence of random partitions and changepoints explicitly.
Let $P_{c^{(z)}_{r, t}, k}$ be the probability that measurement $r$ is assigned at time $t$ to state $k$ for subjects with profile $z$.
Many alternatives to the tRPM model the evolution of $c^{(z)}_{r, t}$ indirectly by encouraging similarity between $P_{c^{(z)}_{r, t}, k}$ at consecutive time steps \citep[e.g.][]{caron_generalized_2017, nieto-barajas_time-series_2012}.
In comparison, partitions produced  the tRPM can exhibit a stronger correlation and evolve smoothly over time, which is preferable when estimating the partition is of direct interest \citep{page_dependent_2022}.
Unlike the original formulation of the tRPM, that at each time step resamples the likelihood parameters corresponding to partition sets, we let states be potentially shared both across profiles and time steps.
This choice helps with their estimation, as demonstrated in our simulations, and allows us to learn highly interpretable measurement clusters that capture associations between measurements at the same as well as at different time steps. 

\subsection{Robust likelihood model} 

In our temporal biclustering model, all observations from measurements assigned to state $k$ share the same likelihood model:
\begin{equation}
    Y_{i, r, t} \mid c^{(s_i)}_{r,
    t} = k, \theta^*_k \iid F_{\boldsymbol{\theta}^*_k}.
\end{equation}
\noindent While this greatly benefits the interpretability of states and their posterior estimation, to make this assumption realistic we want to choose a likelihood model that is diffuse enough to allow for some degree of heterogeneity between observations assigned to the same state.
In our applications, we let $F_{\boldsymbol{\theta}^*_k}$ be a location-scale t-distribution with small degrees of freedom $\nu$, and we let $\boldsymbol{\theta}^*_k = (\mu_k, \sigma^2_k)$.
The appeal of this choice of likelihood becomes especially clear in light of its interpretation as a mixture of normal distributions with common location and variances distributed according to a scaled inverse chi-squared distribution:
%
\begin{align}
    Y_{i, r, t} \mid c^{(s_i)}_{r, t} = k, \mu_k, V_{i, r, t} &\ind \text{Normal}(\mu_k, V_{i, r, t}), \label{eq:lik_nor} \\
    V_{i, r, t} \mid c^{(s_i)}_{r, t} = k, \sigma^2_k &\iid \text{Inv-}\chi^2(\nu, \sigma^2_k). \label{eq:lik_chi}
\end{align}
Eqs.~\eqref{eq:lik_nor} and \eqref{eq:lik_chi} imply that observations from measurements assigned to the same state have the same expected value, the departure from which can occasionally be substantial depending on the level of noise specific to an observation.
We complete this choice of likelihood by assuming that the location and scale parameters are a-priori independently and identically distributed from, respectively, a Normal and a Gamma hyperprior. 

\subsection{Priors on cluster assignments} \label{subsec:cluster_prior}

\begin{figure}[!t]
\centering
\resizebox{.7\textwidth}{!}{
    \begin{tikzpicture}[roundnode/.style={circle, draw=black}]
    
        \node[roundnode, scale=1.2, fill=lightgray](Y_r1i) {$Y_{i, r, 1}$};  
        \node[roundnode, right=1 of Y_r1i, scale=1.2, fill=lightgray](Y_r2i) {$Y_{i, r, 2}$};    
        \node[roundnode, right=1 of Y_r2i, scale=2.1, fill=lightgray](Y_dot) {$\dots$};
        \node[roundnode, right=1 of Y_dot, scale=1.2, fill=lightgray](Y_rTi) {$Y_{i, r, T}$};     
        \node[roundnode, above left= 0.5 and 3.5 of Y_r1i, scale=1.9, fill=white](s_i) {$s_i$}; 
        
        \draw[->, >=stealth'] (s_i.east) to [out=0,in=90, looseness=0.65] ($(Y_r1i.west)!0.8!(Y_r1i.north west)$);  
        \draw[->, >=stealth'] (s_i.east) to [out=0,in=90, looseness=0.5] ($(Y_r2i.west)!0.8!(Y_r2i.north west)$);   
        \draw[->, >=stealth'] (s_i.east) to [out=0,in=90, looseness=0.4] ($(Y_dot.west)!0.8!(Y_dot.north west)$);
        \draw[->, >=stealth'] (s_i.east) to [out=0,in=90, looseness=0.3] ($(Y_rTi.west)!0.8!(Y_rTi.north west)$);
        
        \node[roundnode, left=1.4 of s_i, scale=2.1](pi) {$\pi$};     
        \node[roundnode, above left=0.2 and 1.2 of pi, scale=1.8](alpha) {$\zeta$};
        \node [draw, diamond, aspect=1, below left=0.2 and 1.2 of pi, scale = 1.2, fill=lightyellow](Z) {$Z$};
        \node [draw, diamond, aspect=1,  above left=0.2 and 1.2 of alpha, scale = 1.1, fill=lightyellow](b_1) {$b_1$};
        \node [draw, diamond, aspect=1,  below left=0.2 and 1.2 of alpha, scale = 1.1, fill=lightyellow](b_2) {$b_2$};        
    
        \draw[->, >=stealth'] (pi) -- (s_i);    

        \draw[->, >=stealth'] (alpha) -- (pi);  
        \draw[->, >=stealth'] (Z) -- (pi);
        \draw[->, >=stealth'] (b_1) -- (alpha);
        \draw[->, >=stealth'] (b_2) -- (alpha);        
        
        \node[roundnode, above=1.5 of s_i, scale=1.6, fill=white](theta_k) {$\theta_k$};

        \node [draw, diamond, aspect=1, left=1 of theta_k, scale = 1.2, fill=lightyellow](theta_0) {$\theta_0$};    

        \draw[->, >=stealth'] (theta_0) -- (theta_k);   

        \draw[->, >=stealth'] (theta_k.east) to [out=0,in=90, looseness=1.8] ($(Y_r1i.north west)!0.6!(Y_r1i.north)$);
        \draw[->, >=stealth'] (theta_k.east) to [out=0,in=90, looseness=1.45] ($(Y_r2i.north west)!0.6!(Y_r2i.north)$);
        \draw[->, >=stealth'] (theta_k.east) to [out=0,in=90, looseness=1.25] ($(Y_dot.north west)!0.6!(Y_dot.north)$);
        \draw[->, >=stealth'] (theta_k.east) to [out=0,in=90, looseness=1.05] ($(Y_rTi.north west)!0.6!(Y_rTi.north)$); 

        \node[roundnode, above=5.5 of Y_r1i, scale=1.3, fill=white](c_r1z) {$c_{r,1}^{(z)}$};    
        \node[roundnode, above=5.5 of Y_r2i, scale=1.3, fill=white](c_r2z) {$c_{r,2}^{(z)}$};    
        \node[roundnode, above=5.5 of Y_dot, scale=2.1, fill=white](c_dot) {$\dots$};
        \node[roundnode, above=5.5 of Y_rTi, scale=1.3, fill=white](c_rTz) {$c_{r,T}^{(z)}$};
        
        \draw[->, >=stealth'] (c_r1z) -- (c_r2z);   
        \draw[->, >=stealth'] (c_r2z) -- (c_dot);
        \draw[->, >=stealth'] (c_dot) -- (c_rTz);

        \draw[->, >=stealth'] (c_r1z) -- (Y_r1i);   
        \draw[->, >=stealth'] (c_r2z) -- (Y_r2i);
        \draw[->, >=stealth'] (c_dot) -- (Y_dot);
        \draw[->, >=stealth'] (c_rTz) -- (Y_rTi);
        
        \node[roundnode, above=4.5 of pi, scale=1.7, fill=white](omega_0) {$\omega_{0}$};       
        \node[roundnode, above=5.5 of s_i, scale=1.3, fill=white](omega_z) {$\omega^{(z)}$};  
        \node[roundnode, above left=0.2 and 1 of omega_0, scale=2, fill=white](alpha_0) {$\eta$};  
        \node [draw, diamond, aspect=1, below left=0.2 and 1 of omega_0, scale = 1.2, fill=lightyellow](K) {$K$};
        \node [draw, diamond, aspect=1, above left=0.2 and 1 of omega_z, scale = 1.3, fill=lightyellow](alpha_omega) {$\phi$};
        \node [draw, diamond, aspect=1, above left=0.2 and 1 of alpha_0, scale = 1.1, fill=lightyellow](c_omega) {$d_1$};    
        \node [draw, diamond, aspect=1, below left=0.2 and 1 of alpha_0, scale = 1.1, fill=lightyellow](d_omega) {$d_2$};                     
        
        \draw[->, >=stealth'] (alpha_0) -- (omega_0);   
        \draw[->, >=stealth'] (K) -- (omega_0); 
        \draw[->, >=stealth'] (omega_0) -- (omega_z);        
        \draw[->, >=stealth'] (alpha_omega) -- (omega_z);
        \draw[->, >=stealth'] (c_omega) -- (alpha_0);
        \draw[->, >=stealth'] (d_omega) -- (alpha_0);        

        \draw[->, >=stealth'] (omega_z.east) to [out=0,in=90, looseness=0.65] ($(c_r1z.north west)!0.6!(c_r1z.north)$);
        \draw[->, >=stealth'] (omega_z.east) to [out=0,in=90, looseness=0.5] ($(c_r2z.north west)!0.6!(c_r2z.north)$);
        \draw[->, >=stealth'] (omega_z.east) to [out=0,in=90, looseness=0.4] ($(c_dot.north west)!0.6!(c_dot.north)$);
        \draw[->, >=stealth'] (omega_z.east) to [out=0,in=90, looseness=0.3] ($(c_rTz.north west)!0.6!(c_rTz.north)$);
        
        \node[roundnode, above=1.8 of c_r2z, scale=1.3, fill=white](gamma_r2) {$\gamma_{r,2}^{(z)}$};    
        \node[roundnode, above=1.8 of c_dot, scale=2.1, fill=white](gamma_dot) {$\dots$};
        \node[roundnode, above=1.8 of c_rTz, scale=1.3, fill=white](gamma_rT) {$\gamma_{r,T}^{(z)}$};     
        
        \draw[->, >=stealth'] (gamma_r2) -- (c_r2z);    
        \draw[->, >=stealth'] (gamma_dot) -- (c_dot);
        \draw[->, >=stealth'] (gamma_rT) -- (c_rTz);    

        \node[roundnode, above=1 of gamma_r2, scale=1.3, fill=white](a_2) {$a_{2}^{(z)}$};  
        \node[roundnode, above=1 of gamma_dot, scale=2.1, fill=white](a_dot) {$\dots$};
        \node[roundnode, above=1 of gamma_rT, scale=1.3, fill=white](a_T) {$a_{T}^{(z)}$};
        \node [draw, diamond, aspect=1, above left=1.1 and 0.2 of a_dot, scale = 1.4, fill=lightyellow](alpha_a) {$\alpha$};
        \node [draw, diamond, aspect=1, above right=1.1 and 0.2 of a_dot, scale = 1.2, fill=lightyellow](beta_a) {$\beta$};
        
        \draw[->, >=stealth'] (alpha_a) -- (a_2);
        \draw[->, >=stealth'] (alpha_a) -- (a_dot);
        \draw[->, >=stealth'] (alpha_a) -- (a_T);
        \draw[->, >=stealth'] (beta_a) -- (a_2);
        \draw[->, >=stealth'] (beta_a) -- (a_dot);
        \draw[->, >=stealth'] (beta_a) -- (a_T);
        
        \draw[->, >=stealth'] (a_2) -- (gamma_r2);  
        \draw[->, >=stealth'] (a_dot) -- (gamma_dot);
        \draw[->, >=stealth'] (a_T) -- (gamma_rT);   
        
        \begin{scope}[on background layer]
        \tikzset{plate caption/.append style={below=20pt of #1.south east}}
        \plate[rounded corners=0pt, inner sep=0.5cm, xshift=-0.1cm, yshift=0.1cm, color=black, fill=col_plate_R, fill opacity=0.15] {plate_R} {(Y_r1i) (Y_r2i) (Y_dot) (Y_rTi) (c_r1z) (c_r2z) (c_dot) (c_rTz) (gamma_r2) (gamma_dot) (gamma_rT)} {$R$}; 
        \end{scope} 

        \begin{scope}[on background layer]
        \tikzset{plate caption/.append style={below=2pt of #1.south east}}
        \plate[rounded corners=0pt, inner sep=0.25cm, xshift=-0.1cm, yshift=0cm, color=black, fill=col_plate_Z, fill opacity=0.1] {plate_Z} {(c_r1z) (c_r2z) (c_dot) (c_rTz) (omega_z) (gamma_r2) (gamma_dot) (gamma_rT) (a_2) (a_dot) (a_T)} {$Z$}; 
        \end{scope} 

        \begin{scope}[on background layer]
        \tikzset{plate caption/.append style={below=2pt of #1.south east}}
        \plate[rounded corners=0pt, inner sep=0.2cm, xshift=-0.1cm, yshift=0cm, color=black, fill=col_plate_N, fill opacity=0.1] {plate_N} {(Y_r1i) (Y_r2i) (Y_dot) (Y_rTi) (s_i)} {$N$}; 
        \end{scope} 

        \begin{scope}[on background layer]
        \tikzset{plate caption/.append style={below=2pt of #1.south east}}
        \plate[rounded corners=0pt, inner sep=0.15cm, xshift=-0.1cm, yshift=0cm, color=black, fill=col_plate_mu, fill opacity=0.1] {plate_theta} {(theta_k)} {$K$}; 
        \end{scope} 
        
        
     \end{tikzpicture}
     }
     \caption{\small Directed graphical representation of our MMTB model. The likelihood of observations $Y_{i,r,t}$ at time $t = 1, \dots, T$ depends on the profile $s_i$ of subject $i = 1, \dots, N$. When $s_i = z$ and the state $c_{r,t}^{(z)}$ of measurement $r$ equals $k$, then $Y_{i,r,t}$ has a distribution $F_{\boldsymbol{\theta}_k}$, for $r  \in 1, \dots, R$ and $k \in \{1, \dots, K\}$. $\boldsymbol{\theta}_0$ parameterize the hyperprior on $\boldsymbol{\theta}_k$. Profiles $s_i$ are sampled from a categorical distribution with probabilities $\boldsymbol{\pi} = (\pi_1, \dots, \pi_Z)$. For subjects with profile $s_i = z$, the state of measurement $r$ at time $t$ is the same as at $t-1$ when the persistence indicator $\gamma_{r,t}^{(z)}$ equals $1$, which has probability $a_t^{(z)}$. Otherwise, the state is re-sampled from a categorical distribution with probabilities $\boldsymbol{\omega}^{(z)}= (\omega^{(z)}_{1}, \dots, \omega^{(z)}_{K})$.  Yellow, diamond-shaped nodes denote hyperparameters with pre-specified values.}
  \label{fig:graphical_model}
\end{figure}
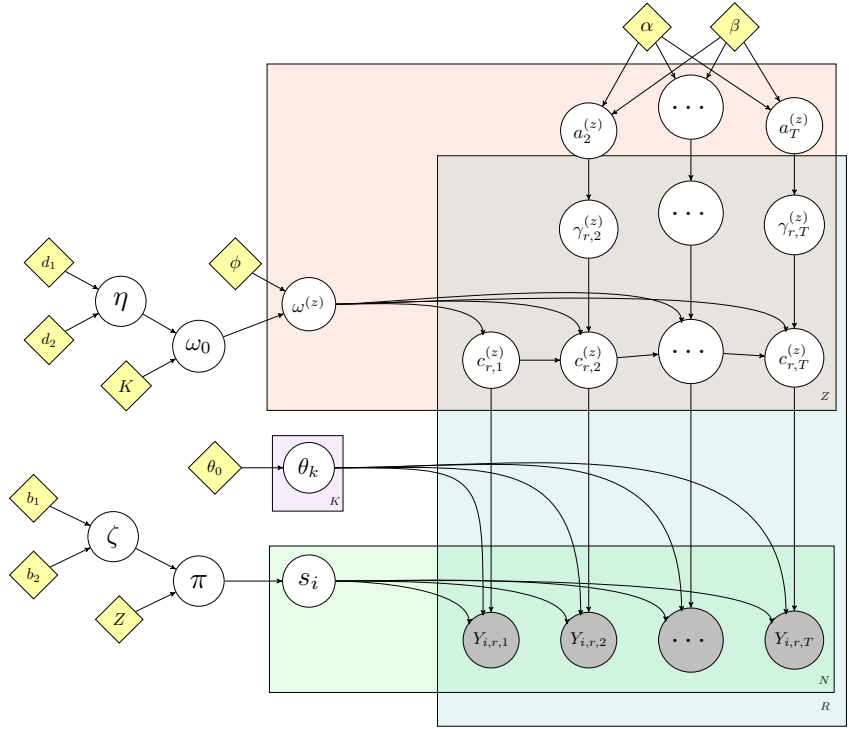

We are left to assign a prior distribution for the assignment variables and the probabilities of profiles and states (that is, respectively, subject and measurement clusters).
In both cases, we opt for the sparse finite mixture approach by \cite{malsiner-walli_model-based_2016}, which is a tractable yet flexible prior that allows us to estimate the number of active clusters from data. 
In particular, we let the prior on profile assignments and their probabilities  $\boldsymbol{\pi} =  (\pi_1, \dots, \pi_Z) $ be
\begin{align}
    s_i \mid \boldsymbol{\pi} &\iid \text{Categorical}\left(\pi_1, \dots, \pi_Z \right), \label{eq:profile_prior} \\
    \boldsymbol{\pi} \mid \zeta &\sim \text{Dirichlet} \left(\dfrac{\zeta}{Z}, \dots, \dfrac{\zeta}{Z} \right), \label{eq:profile_prob_prior}
\end{align}
where we let $Z = N$, and place a hyperprior $\zeta \sim \text{Gamma}(b_1, b_2)$ with $b_1$ and $b_2$ favoring small values of $\zeta$.
When supported by the data, the posterior of $\boldsymbol{\pi}$ can concentrate on vectors assigning all but negligible probability to a subset of the $Z$ profile indices.
Further, we place a hierarchical Dirichlet prior on profile-specific categorical distributions over the $K$ states:
\begin{align}
    \boldsymbol{\omega}^{(z)} \mid \boldsymbol{\omega}_0 &\iid \text{Dirichlet}(\phi \, \omega_{01}, \dots, \phi \, \omega_{0K}), \quad z = 1, \dots, Z \\
    \boldsymbol{\omega}_0 \mid \eta &\sim \text{Dirichlet}\left(\dfrac{\eta}{K}, \dots, \dfrac{\eta}{K}\right), 
\end{align}
where similar to profile probabilities, we let $\boldsymbol{\omega}_0 = (\omega_{01}, \dots, \omega_{0K})$ and $ \eta \sim \text{Gamma}(d_1, d_2)$.
When $K$ is large and $\eta \ll K$, this prior distribution approximates the hierarchical Dirichlet Process \citep{ishwaran_dirichlet_2002, green_modelling_2001, teh_hierarchical_2006}, whose predictive distributions are described via the so-called Chinese restaurant franchise (CRF). 
The posteriors of $\boldsymbol{\omega}_0$ and $\boldsymbol{\omega}^{(z)}$ can assign all but negligible mass to a subset of the $K$ available states, and therefore we can learn a number of active states smaller than $K$ from the data \citep{malsiner-walli_model-based_2016}.
%
The proposed model is summarized in Figure \ref{fig:graphical_model} via a graphical model encoding the conditional dependencies between variables. 

\hypertarget{sec:posterior}{%
\section{Posterior inference}\label{sec:posterior}}

We propose a method for efficiently fitting our model to data to obtain the posterior distribution of the variables of interest, devoting special attention to the derivation of our efficient blocked update of state sequences. Then, we describe how MCMC samples can be aggregated to summarize posterior estimates of profiles and changepoints.

\subsection{MCMC overview}
The marginal posterior distribution of the variables of main interest --- dynamic connectivity profiles and corresponding state sequences for each measurement, state-specific likelihood parameters and assignment of subjects to profiles --- cannot be derived in closed form.
Hence, we fit the model by exploiting the conditional independencies illustrated in Figure \ref{fig:graphical_model}, using conjugate priors where appropriate to devise a simple-to-implement and efficient Gibbs sampling algorithm structured as follows:
%

\begin{enumerate}
    \item Update variables related to the clustering of subjects into profiles: \label{alg:update_profiles}
    \begin{enumerate}
        \item[-] Sample profile probabilities $\boldsymbol{\pi}$ conditional on subject assignments and resample their concentration hyperparameter $\zeta$ using a Metropolis-Hastings (MH) update as in Algorithm 2 in \cite{fruhwirth-schnatter_here_2019}; \label{alg:update_profile_conc}
        \item[-] Sample subjects' assignment to profiles $(s_1, \dots, s_N)$ conditional on $\boldsymbol{\pi}$.
    \end{enumerate}
    \item Update variables related to the temporal clustering of measurements into states:
    \begin{enumerate}
        \item[-] Sample the vector of global state probabilities $\boldsymbol{\omega}_0$ using auxiliary-variable methods developed for the CRF \citep{teh_hierarchical_2006}, and its concentration hyperparameter $\eta$ using a Metropolis Hastings (MH) as in Step \ref{alg:update_profile_conc};\label{alg:update_omega0}
        \item[-] For each profile $z$, sample the profile-specific vector of state probabilities $\boldsymbol{\omega}_z$ conditional on $\boldsymbol{\omega}_0$; \label{alg:update_omegak}
        \item[-] For each profile $z$, measurement $r$, and time step $t$, sample the state persistence indicator $\gamma^{(z)}_{r, t}$ and the state assignment $c^{(z)}_{r, t}$ as detailed is Section \ref{subsec:block}; \label{alg:update_c}
        \item[-] For each profile $z$ and time step $t$, sample the probability of state persistence $a^{(z)}_{t}$ conditional on all state persistence indicators. \label{alg:update_alpha}
    \end{enumerate}    
    \item Update likelihood parameters: \label{alg:update_lik}
    \begin{enumerate}
        \item[-] For each state $k$, sample its associated likelihood parameters $\boldsymbol{\theta}_k$ conditional on state assignment sequences for all observations.
    \end{enumerate}
\end{enumerate}

\subsection{Block-update of measurement partition and changepoints} \label{subsec:block}
While most steps of our MCMC can efficiently be performed combining existing techniques, we derive a novel blocked sample update of state assignments and persistence indicators' sequences to implement step \ref{alg:update_c}.
For this step in the single-subject scenario, \cite{page_dependent_2022} devise marginal updates, where a state assignment is resampled conditional on the sequence of state persistence indicators and on \textit{both} previous and subsequent state assignments.
With this method, at each MCMC iteration, changes in a state assignment can only be considered if they are compatible with all current values of state assignments and persistence indicators.
This can lead to very slow exploration of the posterior, as is easily seen from a concrete example.
Let $t$ and $r$ be a given time step and measurement, and suppose that the sequence of state assignments and state persistence indicators in a small window around time $t$ at iteration $m$ are, respectively, $(c^{(m)}_{r, t-2}, c^{(m)}_{r, t-1}, c^{(m)}_{r, t}, c^{(m)}_{r, t+1}) = (1, 2, 2, 2)$ and $(\gamma^{(m)}_{r, t-2}, \gamma^{(m)}_{r, t-1}, \gamma^{(m)}_{r, t}, \gamma^{(m)}_{r, t+1}) = (1, 0, 1, 1)$.
If the sequence of state assignments $(c_{r, t-2}, c_{r, t-2}, c_{r, t}, c_{r, t+1}) = (1, 1, 1, 1)$ had higher posterior probability, it would take several MCMC iterations to reach this configuration.
Indeed, even though $\gamma^{(m)}_{r, t-1} = 0$, directly changing $c^{(m)}_{r, t-1}$ from $2$ to $1$ is not possible because $\gamma^{(m)}_{r, t} = 1$ imposes that the state at time $t-1$ must equal the state at time $t$ which currently is $c^{(m)}_{r, t} = 2$.
As suggested by this simple example, the marginal sampler can struggle to explore the posterior, especially for long sequences with substantial temporal dependence. 

\begin{figure}[t]
    \centering
    \includegraphics{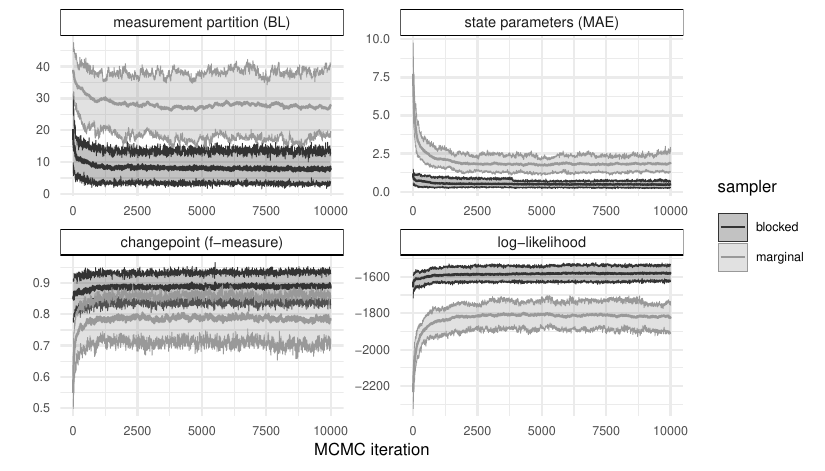}
    \caption{Performance of MCMC with blocked versus marginal updates of state and persistence indicator sequences. Subplots show average and 90\% credible intervals across 30 simulated datasets featuring both time and subject dependence. Simulation details are detailed in Sec. \ref{sec:simulations} (third simulation scenario). Lower values of binder loss (BL) and mean absolute error (MAE) are preferred; higher values of log-likelihood and f-measure are preferred. }
    \label{fig:block_vs_marginal}
\end{figure}

In an experiment on 30 simulated datasets, the marginal sampler systematically failed to reach most-likely configurations of posterior parameters even after 10,000 iterations, as shown in Figure \ref{fig:block_vs_marginal}.  
Posterior samples visited by the blocked sampler, that was initialized at the same random parameter configuration as the marginal sampler, consistently achieved lower binder loss (BL) between sampled and true measurement partitions, and lower mean absolute error (MAE) between sampled and true state-specific location parameters. 
Further, larger f-measures indicate more accurate recovery of changepoints in the state sequences, and larger log-likelihoods suggest that the posterior samples from the blocked sampler better fit the data.
Details on computation of posterior summaries and BL are provided in Sec.~\ref{subsec:posterior_summaries}.

Potential issues with marginal samplers were previously noted by \cite{scott2002bayesian} and by \cite{fox_sticky_2011}, who derived a blocked sampler for the Hierarchical Dirichlet Process Hidden Markov Model and empirically verified its superior performance to the marginal sampler.
Likewise, we derive a variant of the forward-backward algorithm \citep{rabiner_tutorial_1989} for a blocked update of state assignment and persistence indicator sequences in the tRPM.
Conditional on subjects' profile assignments $(s_1, \dots, s_N)$, for each profile $z$ and measurement $r$, we wish to update the sequence of state assignments $\left(c^{(z)}_{r, 1}, \dots, c^{(z)}_{r, T}\right)$ in block, that is by updating the sequence altogether from time $1$ to time $T$. 
In practice, we can achieve this by updating both $c^{(z)}_{r, t}$ and  $\gamma^{(z)}_{r, t}$ conditionally only on $c^{(z)}_{r, t-1}$, rather than also on $c^{(z)}_{r, t+1}$ and  $\gamma^{(z)}_{r, t+1}$, as in a marginal sampler.
To do so, we need to first compute messages backward, which will allow us to marginalize over future state assignments by directly accounting for the likelihood of observations at later times when sampling the sequence in block forward.
This adds some computational cost but enables the sampler to make substantial changes to state sequences over consecutive iterations. For a given profile $z$ and measurement $r$, let $m_{k, t}^{(z, r)}$ be the backward message for state $k$ at time $t$
representing how likely observations \textit{after} time $t$ are if at time $t$ we were in state $k$. 
To compute messages, we start by computing a $K \times T$ matrix of likelihoods, an element of which represents the likelihood of observations of measurement $r$ from subjects with profile $z$ if the state of measurement $r$ at time $t$ was $k$:
$$ \mathcal{L}_{k, t}^{(z, r)}(Y) = \prod_{i: s_i = z} F(Y_{i, r, t} \mid \theta_k). $$
Then, to compute the messages, we start from the last time step by setting $m_{k, T}^{(z, r)} = 1$ for all $k = 1, \dots, K$, since there are no observations after time $T$ and so all possible states are equally likely from the future's perspective. 
We compute the remaining messages backward for $t = T-1, \dots, 1$, recalling that at time $t + 1$ we stay in the state assigned at time $t$ with probability $a_{t+1}^{(z)}$ or otherwise we sample a state $h \in \{1, \dots, K \}$ with probability $\omega^{(z)}_h$. 
Thus,
\begin{equation}
    m_{k, t}^{(z, r)} = a_{t+1}^{(z)} \mathcal{L}_{k, t+1}^{(z, r)}(Y) \, m_{k, t+1}^{(z, r)} + \left(1 - a_{t+1}^{(z)}\right) \sum_{h = 1}^{K} \, \omega^{(z)}_h \, \mathcal{L}_{h, t+1}^{(z, r)}(Y) \, m_{h, t+1}^{(z, r)}. \label{eq:message}
\end{equation}
Intuitively, Eq.~\eqref{eq:message} establishes that, when computing the message to state $k$, we weight the likelihood of future observations --- given by the likelihood of the next time step and the message from later time steps --- by the probability of staying in state $k$ or of sampling a state $h \in \{1, \dots, K \}$.
Once all messages have been computed, we assign time $1$ to state $k$ according to $P(c^{(z)}_{r, 1} = k) \propto \omega_k  \mathcal{L}_{k, 1}^{(z, r)}(Y) m_{k, 1}^{(z, r)}$ and, for $t = 2, \dots, T$, we recursively simulate the sequence of state assignment and state persistence indicators  as follows:
\begin{align}
  P\left(c_{r,t}^{(z)} = c_{r, t-1}^{(z)} = k, \gamma_{r,t}^{(z)} = 1 \right)  &\propto a_t^{(z)} \mathcal{L}_{k, t}^{(z, r)}(Y) m_{k, t}^{(z, r)}, \\[10pt]     
  P\left(c_{r,t}^{(z)} = h,  \gamma_{r,t}^{(z)} = 0 \right) &\propto \left(1 - a_t^{(z)}\right) \omega_{h}^{(z)} \mathcal{L}_{h, t}^{(z, r)}(Y) m_{h, t}^{(z, r)}, \quad h = 1, \dots, K. \label{eq:block_update}
\end{align}
Note that via Eq.~\eqref{eq:block_update}, we simultaneously sample whether to stay in the same state as at the previous time step ($\gamma_{r,t}^{(z)} = 1$), or whether to resample a state ($\gamma_{r,t}^{(z)} = 0$) according to $\omega_{h}^{(z)}$, as well as the state resampled.
For each of the $Z$ profiles and $R$ locations, the block update of states for all times requires $\mathcal{O}(K T)$ computations, faster than the $\mathcal{O}(K^2 T)$ cost of blocked resampling for general hidden Markov models~\citep{scott2002bayesian,fox_sticky_2011}.  We can reduce time substantially by resampling the $Z$ profiles and $R$ locations in parallel.

\subsection{Posterior summaries} \label{subsec:posterior_summaries}

Posterior samples obtained via MCMC chain(s) can be combined into estimates of subjects' profiles and measurements' state sequences.
With respect to profiles, we compute the pairwise posterior probability of co-clustering $P(s_i = s_j)$ for any pair of subjects $i$ and $j$ as the proportion of posterior samples where $i$ and $j$ had the same profile assignment.
We also derive a point estimate of profile assignments based on the Binder loss (BL) function \citep{binder_bayesian_1978}.
The BL of estimating partition $\mathbf{s}^* = (s^*_1, \dots, s^*_N)$ with partition $\mathbf{s} = (s_1, \dots, s_N)$ is defined as $B(\mathbf{s}^*, \mathbf{s}) = \sum_{i < j} a \mathbbm{1}(s^*_i = s^*_j)\mathbbm{1}(s_i \neq s_j) + b \mathbbm{1}(s^*_i \neq s^*_j)\mathbbm{1}(s_i = s_j)$, where $a$ and $b$ are two positive parameters that represent the cost of, respectively, failing to cluster together items that should in fact be together versus clustering together items that should in fact be separated. 
We use the R package \textit{salso}~\citep{dahl_search_2022} to find the partition $\tilde{\mathbf{s}}$ that minimizes the expected Binder loss $E[B(\mathbf{s}^*, \mathbf{s})] \approx \frac{1}{M} \text{argmin}_{\mathbf{s}} \sum_{m = 1}^{M} B(\mathbf{s}^{(m)}, \mathbf{s})$ estimated from $M$ posterior samples of state assignment vectors $(\mathbf{s}^{(1)}, \dots, \mathbf{s}^{(M)})$.
By combining estimated pairwise subject co-clustering probabilities in a $N \times N$ symmetric association matrix, sorting subjects according to their estimated profile, we can visualize the estimated subject partition, highlighted by diagonal blocks of high co-clustering probabilities, as well as uncertainty in profile assignments.
To summarize results with respect to state sequences, we consider the estimated partition of subjects into profiles and, for each estimated profile, obtain profile-specific sequences of location parameters by averaging the state-specific location parameter sampled at any given time step across all subjects assigned to that profile and across all samples and MCMC chains. 
In a similar way, we can derive estimated, profile-specific sequences of changepoint probabilities by averaging state-persistence indicator sequences of all subjects assigned to a profile.

\hypertarget{sec:applications}{%
\section{Application to neuroscience studies}\label{sec:applications}}

\begin{figure}[t]
    \centering
    \includegraphics[width = 0.95\textwidth]{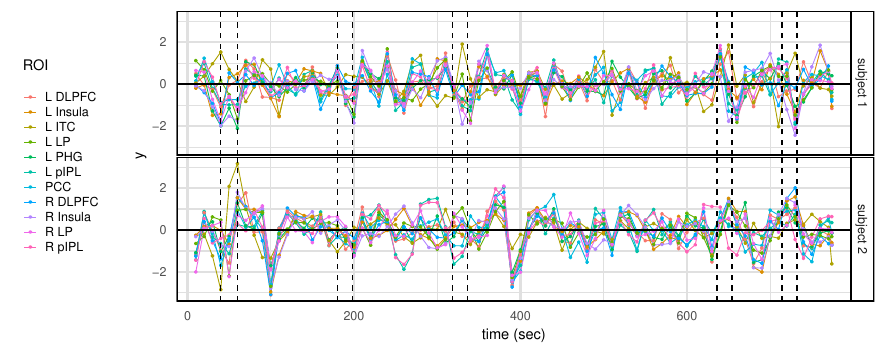}
    \caption{fMRI data after pre-processing for two subjects and all selected brain regions of interest (ROIs). The vertical axis plots BOLD signal averaged across voxels in a ROI over a 10-second window. Vertical dashed lines demark five activation blocks where subjects squeeze a ball.}
    \label{fig:fmri_data}
\end{figure}

The temporal clustering framework we propose in this work can be employed in a variety of studies where multiple subjects undergo the same experiment and multiple measurements are recorded for each subject at the same time steps.
To illustrate the versatility of the proposed approach, we analyze data collected in two different studies of brain activity: an fMRI and an EEG experiment. 
When analyzing data from multi-subject neuroscience experiments, it is expected that different subjects may exhibit different patterns of neural responses to the same stimuli or task instructions. Current knowledge about the source and mechanisms underlying such individual differences varies widely depending on context or task, psychological or neurological condition, or even sociocultural factors. It is therefore of great interest to develop methods to understand the landscape of such multidimensional individual differences, as is common in fields such as personalized medicine or computational psychiatry. Our MMTB model is specifically designed to identify potential sub-populations within a given cohort based on neural profiles.

\subsection{Parameter settings}\label{hyperparameters-fmri-eeg}
Before delving into each application, we describe the choice of hyperparameters for clustering-related variables and the MCMC settings, which were common across the two applications. For the analyses of fMRI and EEG data, we let the maximum number of profiles $Z$ match the number of subjects $N$ in the data. We favor coarser partitions of subjects by setting $b_1 = 50$  and $b_2 = 100$. To encourage using a subset of the available states, we let $c_1 = 50, c_2 = 100$ and set $\phi = 0.5$. We encourage smooth changes of states over time by letting $\alpha^{(z)}_t \iid \text{Beta}(10, 2)$ be the hyperprior on state-persistence probabilities. 
In the fMRI application, we initialize state location parameters to equally-spaced numbers between -1.5 and 1.5 with increments of .25, resulting in a maximum of $K = 13$ states. We let $\mu_k \iid \text{Normal}(0, 3)$ and $\sigma^2_k \iid \text{Gamma}(5, 10)$.
To define likelihood hyperparameters for the EEG data, we let $K = 20$, initialize location parameters from a prior $\mu_k \iid \text{Normal}(0, 5)$, and assume $\sigma^2_k \iid \text{Gamma}(1, 1)$.
We combine posterior samples from 3 MCMC chains of 10,000 iterations, after discarding the first 5,000 samples as burn-in. Traceplots suggested fast convergence and indicated that the 3 chains visited configurations of the model with similar posterior probabilities, in both applications (see supplementary material).

\subsection{Estimating brain activity profiles based on fMRI data}\label{fmri-data}
Functional magnetic resonance imaging provides an indirect measure of brain activity through Blood Oxygenation Level Dependent (BOLD) contrasts, a signal measured at volumetric pixels (voxels) that span the brain.
We analyze data from a fMRI study conducted at the Center for Advanced Neuroimaging of the University of California Riverside \citep{hussain_locus_2023}. 
Subjects in this study underwent an experiment in which they alternated between 6 resting-state blocks (of 5, 2, 2, 5, 1 and 1 minute, respectively) and 5 squeeze blocks (each lasting 18 seconds) when they were instructed to bring their dominant hand to their chest while compressing a squeeze-ball at maximum grip strength.
The fMRI data was collected using a 2D echo planar imaging sequence with echo time 32 ms and repetition time of 2 sec.
The dataset that we analyze includes 23 subjects. Data for the analysis were restricted to a time window starting 40 sec before the first active session, and lasting 40 sec after the last session, therefore covering 12 min 54 sec of experiment. 
To minimize the potential impact of slight misalignments between observations on the identification of profiles, we average 5 consecutive observations over a 10 sec window, resulting in $T = 78$ time steps.

For our analysis, we focused on 11 brain regions (regions of interest, or ROIs), identified following the anatomical definition in \cite{hussain_locus_2023}. These were chosen due to having a role in either the Default Mode Network (DMN) or the Salience Network (SN). The DMN has often been associated with resting and inattentiveness, and was chosen due to the resting state nature of sections of the experiment \citep{greicius_default-mode_2004}.
For the DMN, we considered Posterior Cingulate Cortex (PCC), left and right posterior Inferior Parietal Lobule (L pIPL and R pIPL), left and right Dorsolateral Prefrontal Cortex (L DLPFC and R DLPFC), left Parahippocampal Gyrus (L PHG) and left Inferolateral Temporal Cortex (L ITC).
The SN regions left and right Insula (L Insula and R Insula) and left and right Lateral Parietal (L LP and R LP) were selected due to being implicated in allocating response to stimuli, like the squeeze instructions \citep{menon_saliency_2010}.
Figure \ref{fig:fmri_data} displays data for two subjects after pre-processing.

\begin{figure}[t]
    \centering
        \includegraphics[width=0.45\columnwidth]{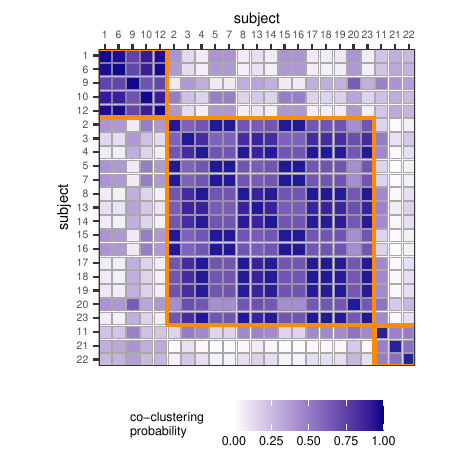}
    \caption{fMRI data: Estimated probability that any pair of subjects be assigned to the same profile. Subjects are sorted by estimated profile memberships, as marked with orange blocks.}
\label{fig:fmri_coclust_blocks}
\end{figure}

\begin{figure}[t]
    \centering
    \includegraphics[width=0.95\columnwidth]{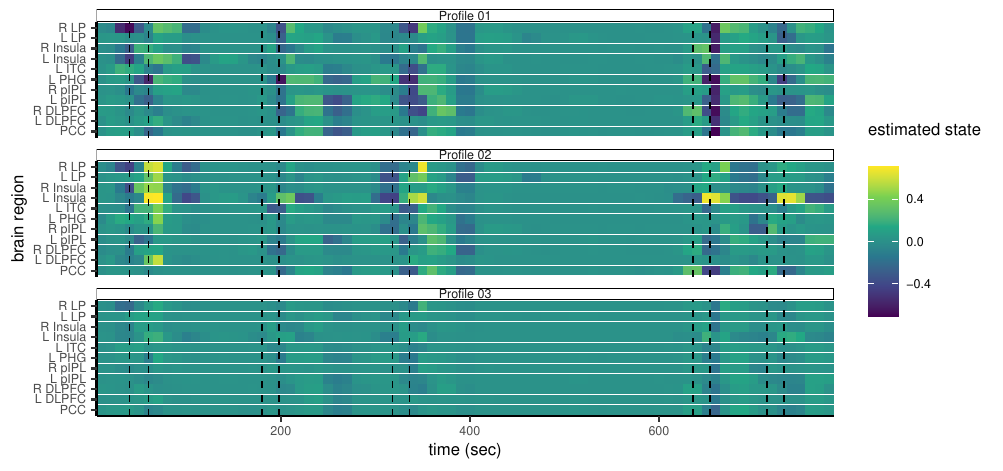}
    \caption{fMRI data: Estimated state-specific location parameters for all selected ROIs and for each of the estimated activity profiles. Vertical dashed lines mark the five squeeze blocks.}    
    \label{fig:fmri_reg_clust}    
\end{figure}

Figure \ref{fig:fmri_coclust_blocks} displays estimated pairwise posterior probabilities of co-clustering among subjects, along with estimated profile assignments. Using the BL to estimate the most likely profile configuration from all posterior samples as detailed in Section \ref{subsec:posterior_summaries}, we estimate the presence of 3 profiles of subjects (activity profiles). Estimated co-clustering probabilities suggest likely overlapping between profiles, particularly between profiles 1 and 3 (enumerating blocks from left to right).
For each estimated profile, Figure \ref{fig:fmri_reg_clust} plots the estimated sequence of location parameters for all ROIs considered, with vertical dashed lines marking the five squeeze blocks during the experiment.
We notice that, for all profiles, state changes are more likely detected in proximity of or during squeeze blocks. Most notably, there is a clustering (co-activation) of R LP and L ITC during the beginning and ending of the squeeze block.
Subjects in profile 2 seem to exhibit more extreme relative brain signal reactions to the beginning and ending of the squeeze block. This appears as a unique increase BOLD signal among the LP, Insula, ITC, and DLPFC at the conclusion of the squeeze block. These ROIs align with the receiving and decoding of the visual cue to put the arm back at rest, switching to the task, and then performing the action. Interestingly, subjects in profile 1 and 3 show similar clustering behavior, but do not exhibit the magnitude of BOLD response as subjects in profile 2.
Right after a squeeze block, profile 1 undergoes a short drop in BOLD signal in L PHG and R LP. Profiles 1 and 3 appear to experience a drop in BOLD signal in almost all ROIs at the end of the fourth squeeze block, which is the first active portion after 5 minutes of inactivity.
Overall, albeit in lower magnitude, Profile 3 appears to display state-changes in the same direction as Profile 1 and this similarity is reflected by the higher co-clustering probability between subjects with these two profiles displayed in Figure \ref{fig:fmri_coclust_blocks}.

\subsection{Neural profiles based on electrophysiological data}
\begin{figure}[t]
    \centering
    \includegraphics[width=0.95\columnwidth]{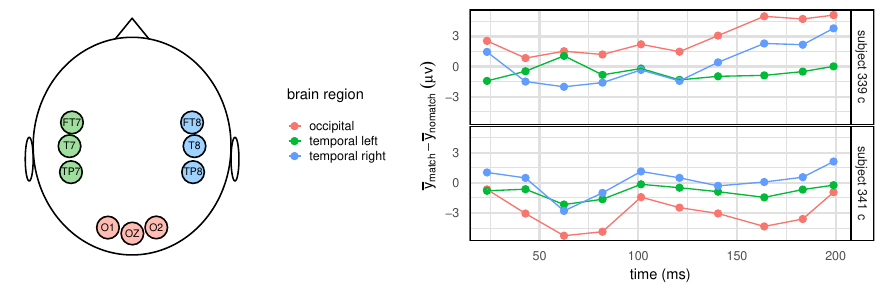}
    \caption{EEG dataset: (Left) Diagram of human head showing relative location of selected electrodes in the occipital, temporal right and temporal left regions. (Right) Data after pre-processing for two subjects. For each region, $\bar{y}_{\text{match}}$ and $\bar{y}_{\text{nomatch}}$ represent the ERP signal in microvolts averaged across selected electrodes and across all trials in which subjects were shown, respectively, two identical or two differing pictures, over a 20 ms window.  }
    \label{fig:eeg_data}
\end{figure}

ERP signals, measuring neural activity in response to specific events (e.g. motor or cognitive stimuli), are
regarded as a powerful, noninvasive way to explore human brain activity \citep{luck_introduction_2014}.
Here we consider a dataset publicly available on the UCI Machine Learning Repository \citep{begleiter_eeg_1999} and first analyzed by \cite{zhang_event_1995} and \cite{ingber_statistical_1998}.
The data was collected during an experiment in which subjects experienced two visual stimuli from the picture set of \cite{snodgrass_standardized_1980}. 
In particular, during a trial, a subject was shown a first picture (S1) for 300 ms and, after 1.6 s, a second picture identical to the first (S2 match) or semantically different (S2 nomatch) for 300 ms.
Each subject completed multiple trials for each condition.
The data includes alcoholic and control subjects, since assessing EEG correlates of genetic predisposition to alcoholism was one of the original purposes of data collection.
More details on experiment and data pre-processing can be found in \cite{zhang_event_1995}.
The available dataset contains ERP measurements in microvolts for 64 electrodes sampled at 256 Hz (3.9-ms epoch). 
We analyze data from all 121 subjects included in the dataset.
For these subjects, we consider all available trials, and we restrict to the first 200 ms recorded during S2 match and S2 nomatch conditions.
To reduce noise, as commonly done with ERP data \citep{luck_introduction_2014}, we average the data recorded across trials so as to obtain a single ERP measurement per subject, electrode position and time step, after stimulus onset, for each of the two conditions.
We then take the difference between the ERP for the S2 match and the S2 nomatch conditions, so as to remove the effect of subject-specific variability (e.g. due to scalp sebum) on the recorded ERP values and we average the values obtained over a 20-ms window, so as to reduce the effect of minimal misalignments between subjects on the identification of subject profiles.
Finally, we averaged data from the electrodes located in the left temporal region, those in the right temporal region and those in the occipital region, obtaining single ERP signals for each subject in these three areas.  These are known to, respectively, affect memory \citep{balderas_consolidation_2008} and be more receptive to visual stimuli \citep{dupont_kinetic_1997}. 
An illustration of selected brain regions and an example of data after pre-processing is shown in Figure \ref{fig:eeg_data}.

\begin{figure}[t]
    \centering
        \includegraphics[width=0.49\textwidth]{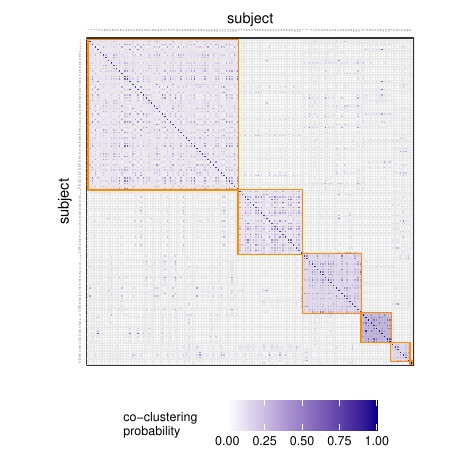}  
    \caption{EEG data: Estimated probability that any pair of subjects be assigned to the same profile. Subjects are sorted by estimated profile memberships, as marked with orange blocks.}
    \label{fig:eeg_coclust_blocks} 
\end{figure}

Figure \ref{fig:eeg_coclust_blocks} displays posterior probabilities of co-clustering among subjects' pairs and estimated profile assignment for each subject. Given the large number of subjects, we consider values for $a$ larger than $b$ in the Binder Loss from Section \ref{subsec:posterior_summaries}. Specifically, for a fixed value $b = 1$, we consider nine equally-spaced values for $a \in [1.1, 1.9]$. Since this determination does not involve re-running the sampler with different hyperparameters, it was simple and only took a few seconds thanks to the function provided in the salso package \citep{dahl_search_2022}. We observed that the estimated partition was fairly robust, as increasing $a$ from $1.1$ to $1.9$ resulted in similar profiles being progressively merged rather than in subjects being moved to different clusters.
By visual inspection of pairwise co-clustering probabilities and state sequences corresponding to the resulting partition of subjects, we found $a = 1.85$ to achieve a good trade-off between interpretability and flexibility, aggregating similar clusters without forcing together subjects with low co-clustering probability (see Figure \ref{fig:eeg_coclust_blocks}).

\begin{figure}[t]
    \centering
    \includegraphics[width=.95\columnwidth]{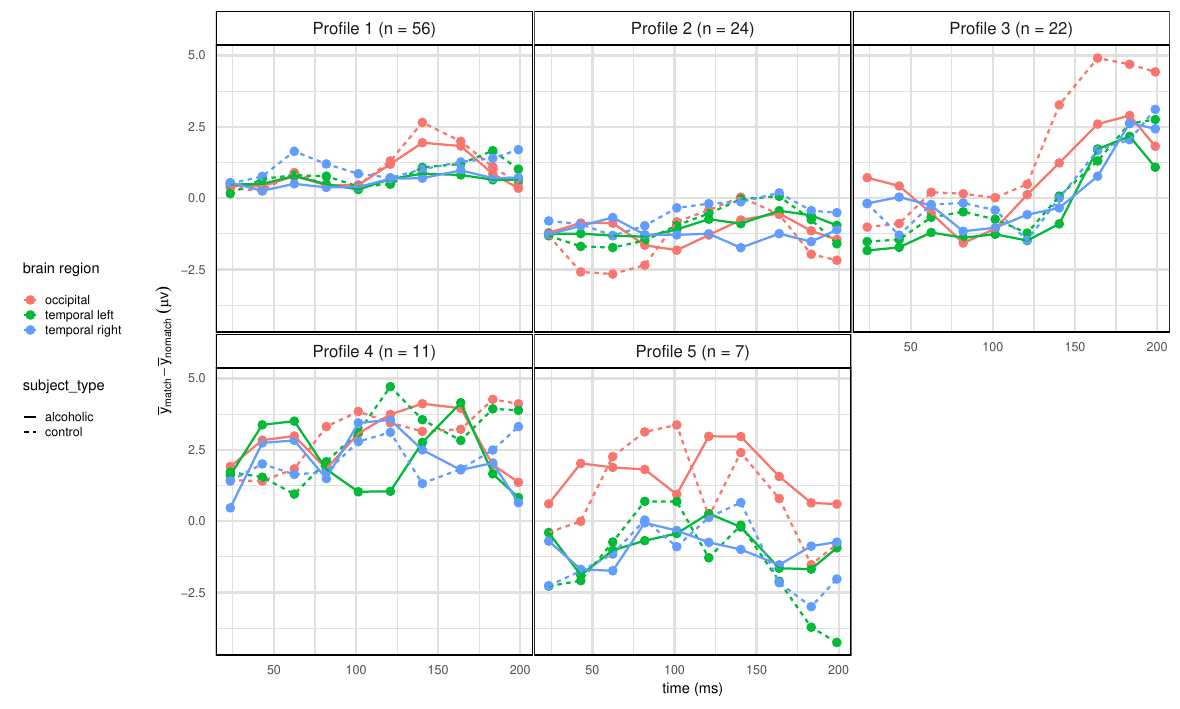}
    \caption{EEG data: Differences in ERP between match and nomatch conditions averaged across subjects with same estimated profile and alcoholic status, for profiles with at least two assigned subjects. The number of subjects assigned to a profile is shown in subplot titles.}    
    \label{fig:eeg_reg_clust}    
\end{figure}

A feature of the proposed temporal biclustering model that emerges from the analysis of the EEG data is the ability to identify temporal patterns shared by several subjects (Profiles 1-5) while identifying subjects that exhibit unique behaviors (Profile 6, again enumerating blocks from left to right in Figure \ref{fig:eeg_coclust_blocks}).
This behavior appears desirable, especially in studies involving many subjects, due to the likely presence of a few outliers.
Recall that the measurements represent the differences in ERP signal under the S2 match and S2 nomatch stimuli. Figure \ref{fig:eeg_reg_clust} shows the observations averaged across subjects within the same estimated profile, for each region and presented separately for alcoholic and control individuals. Overall, subjects with the same estimated profiles are characterized by average ERP differences that mostly exhibit similar trajectories throughout the experiment in both the alcoholic and control groups.
This suggests that the estimated profiles may capture similarities among subjects beyond the observed alcoholic and non-alcoholic characteristics.
Subjects in profiles 1 and 4, on average, exhibited higher ERP signals in the S2 match condition than in the S2 no match condition for the entire 200ms time window. Conversely, subjects with the estimated profile 2 displayed the opposite pattern. 
Furthermore, subjects with profile 3 displayed, on average, small and mostly negative differences between the S2 match and S2 nomatch ERP for the first 125 ms after stimulus onset, followed by relatively large and positive differences from 125 to 200 ms. 
Overall, the signals in the TL and TR regions appear more frequently coupled (in terms of exhibiting similar averages across subjects) compared to the occipital region. Additionally, the occipital region is more likely to show average ERP differences between the two experimental conditions that are larger in magnitude.

\hypertarget{sec:simulations}{%
\section{Simulation Studies}\label{sec:simulations}}

\begin{figure}[t]
    \centering
    \includegraphics{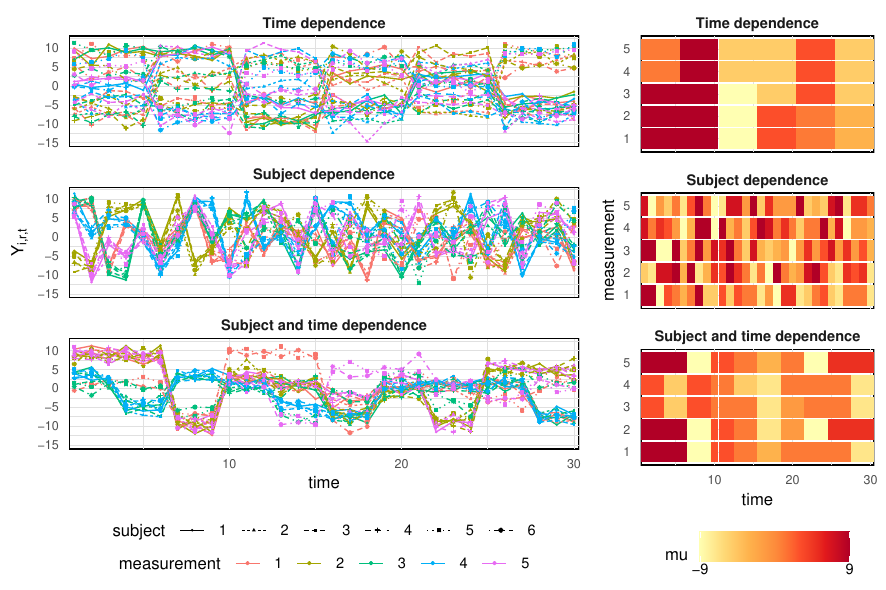}
    \caption{(Left) Simulated data $Y_{i, r, t}$ for subjects $i = 1, \dots, 6$, measurements $r = 1, \dots, 5$, time $t = 1, \dots, 30$, with time or/and subject dependence in state sequences. (Right) Simulated state sequences and corresponding mean parameters for subject $1$ under each simulation scenario.}
    \label{fig:simulation_plots}
\end{figure}

We assess the proposed model and inference methods on simulated data, and also compare the estimation performance against ground-truth values and other methods. 

\subsection{Model comparisons}
To distinguish it from alternative methods, in this section we refer to our model with the acronym {MMTB}, which stands for multi-subject multivariate temporal biclustering.
To illustrate the advantages of simultaneously modeling subjects, measurements and time-point dependencies with our approach, we compare to the following alternative approaches. 
First, we compare with a procedure where our model is applied separately to each subject (single-subject multivariate temporal clustering, {SMTC}), or to each time step (multi-subject multivariate biclustering, {MMB}) or to the mean or median across time for each subject and measurement (respectively, {mean-MMB} and {median-MMB}).
We also compare with the tRPM approach by \cite{page_dependent_2022} (similar to SMTC), that we fit using the  MCMC implementation provided by the authors, and with
the three-dimensional plaid model by \cite{mankad_biclustering_2014} (3d plaid), using the  authors' code.

\subsection{Experimental settings} \label{subsec:simu_data}
We consider three simulation scenarios with data for $N=6$ subjects, $R=5$ measurements and $T = 30$ time steps.
In all scenarios, measurements are partitioned into clusters and their partition can change over time. At every time step, a cluster of measurements is associated with one of $K = 10$ states, corresponding to the parameters used for simulating observations.
The scenarios differ in whether measurement partition and states persist over time, whether measurement partitions are shared across subjects and whether both time and subject dependence are present. 
Details for each scenario are as follows:
\begin{enumerate}
    \item \textbf{Time dependence}: Measurements persist in a state for $5$ consecutive time steps before their state is resampled, and the partition of measurements changes once during the time series. Each subject has different state sequences, i.e. the number of profiles $Z$ is equal to $N$. Simulated data is shown in Figure \ref{fig:simulation_plots} (Top Left) and sequences of states for subject 1 are illustrated in Figure \ref{fig:simulation_plots} (Top Right). 
    \item \textbf{Subject dependence}: The partition of measurements can change at every time step, along with measurements' states. There are two subject profiles: state sequences are shared among subjects $1$, $2$, $3$ and $4$ (Subject $1$), and among subjects $5$ and $6$ (Subject $2$). All simulated data and state sequences for Subject 1 in this scenario are displayed, respectively, in Figure \ref{fig:simulation_plots} (Middle Left) and (Middle Right).
    \item \textbf{Time and subject dependence}: Measurements persist in a state for $3$ consecutive time steps before their state is resampled. The partition of measurements changes twice. Subjects are grouped in the same profiles as in Scenario 2. Figure \ref{fig:simulation_plots} (Bottom Left) and (Bottom Right) show, respectively, all simulated data and state sequences for Subject 1.
\end{enumerate}
%
As the tRPM is implemented with a Normal likelihood, to facilitate comparison we use Normal likelihoods for all simulations and for model fitting; therefore, the 10 states correspond to the mean and variance parameters of the Normal distributions used for simulation.
The 10 means are equally spaced between $-9$ and $9$, and the variances are either $1.5^2$ or $1.25^2$.
We simulate 30 datasets for each scenario, and report average and standard deviation across simulated datasets for all metrics used in our quantitative comparisons. Hyperparameters and other model-fitting settings are described in detail in the supplementary material.

\subsection{Evaluation of subjects' clusters}
Multi-subject models, that is all except SSTC and tRPM, estimate a partition of subjects into clusters (referred to as profiles in this article).
Table \ref{tbl:simu_profiles_results} displays the average BL of estimated versus true subject partitions.
MMTB, mean-MMB, and median-MMB yield a single partition of subjects at every MCMC iteration; we estimate a single partition of subjects from all iterations using the BL (see Sec.~\ref{subsec:posterior_summaries}).
For the MMB model, that estimates a partition of subjects at each time step, we estimate a single partition for each time step combining all MCMC samples and report the average BL across time steps, chains and simulated datasets.
Also for the 3d plaid model, that estimates possibly multiple partitions of subjects, we compute the partition that minimizes the average BL.

The results shown in Table \ref{tbl:simu_profiles_results} indicate a good performance of the proposed MMTB model at recovering the ground-truth subject partition.
MMTB performed very well even if no subject dependence was present, when it correctly identified a different profile for each subject (column 1 in Table \ref{tbl:simu_profiles_results}).
MMB estimates of subject partitions were on average better when this model was fitted independently to each time step than when fitting to the mean or median of observations across time.
The performance of the 3d plaid model at recovering subject partitions was generally lower than the other methods, and we observed a tendency to identify only some of the ground-truth subject clusters.  
Increasing the maximum number of biclusters that the search algorithm was allowed to find did not improve the estimation of bicluster effects for clusters that had not been previously identified.

\subsection{Evaluation of measurements' clusters} \label{subsec:simu_evaluation}
Table \ref{tbl:simu_states_results} reports summaries of models' performance at recovering ground-truth measurement partitions, state-specific location parameters, and changepoints.
Results on measurement partitions are evaluated by computing the average BL (defined in Sec.~\ref{sec:posterior}) between estimated and true measurement partition for each subject and time step.
For the 3d plaid model, we consider all biclusters to which a subject is assigned and estimate a single measurement partition using the BL; as this model does not admit a different partition of measurements per time step, a single one is estimated for the entire time series.
Recovery of likelihood location parameters is assessed with the average absolute mean difference between estimated and true location parameter for each subject and time step; this is  available in the output of the 3d plaid model as the combination of base and biclusters effects.
For methods that model the persistence of measurements' partition over time (the MMTB, SMTC and tRPM) we estimate changepoint probabilities by averaging state persistence indicators across samples and computing the harmonic mean of precision and recall based on those probabilities (f-measure), averaging across subjects and time steps.

As shown in Table \ref{tbl:simu_states_results}, the performance of the MMTB model was consistently good across all metrics and simulation scenarios considered.
Noticeably, MMTB performed well at estimating the sequence of measurement partition, state-specific location parameters and changepoints even when no time dependence was featured in the data (column 6 in Table \ref{tbl:simu_states_results}).
Not surprisingly, treating each subject separately but modeling time dependence with the SSTC or with the tRPM yielded good results when time but not subject dependence was present (first group of columns in Table \ref{tbl:simu_states_results}).
Similarly, when no time dependence was present, estimating likelihood parameters independently for each time step with the MMB let to higher accuracy than all other methods (second group of columns in Table \ref{tbl:simu_states_results}).
Overall, as expected, the tRPM model yielded competitive results for measurement partitions when time-dependence was featured (first and third groups in Table \ref{tbl:simu_states_results}), but proved worse than our model fitted separately to each subject (SSTC) at recovering state parameters and changepoints. 
In particular, the tRPM model exhibited a tendency to overestimate the number of changepoints when state persisted for 5 as well as for 3 consecutive times (respectively, first and third group of columns in Table \ref{tbl:simu_states_results}. This is why, perhaps surprisingly, its results appeared relatively better when simulated state sequences featured subject dependence but low-to-no time dependence (second group of columns in Table \ref{tbl:simu_states_results}).
Finally, the 3d plaid model yielded reasonable results but its performance was generally worse than other methods. 
While worse results on measurement partitions can be attributed to the fact that partition changes over time are not explicitly modeled, the relatively low performance of the 3d plaid model at recovering sequences of mean parameters is more surprising.
Even though this model can in principle, through appropriate time-varying bicluster effects, perfectly fit the sequence of means used for simulation, we observed a tendency to identify only a subset of measurement clusters present in the data. 
We speculate that this performance may partly be due to a combination of the model's flexibility and the greedy approach used by the search algorithm for model fitting.

\begin{table}

\caption{\label{tbl:simu_profiles_results}Binder loss of true vs. estimated subject partition. Displayed are averages (standard deviation) over 30 datasets in simulation scenarios with time and/or subject dependence in state sequences. Lower values are preferred.}
\centering

\begin{tabular}[t]{|l| c c c|c c c|c c c|>{}c|}
\hline
\multicolumn{1}{|c|}{ } & \multicolumn{1}{|p{2.5cm}}{\centering Time dependence} & \multicolumn{1}{p{2.6cm}}{\centering Subject dependence} & \multicolumn{1}{p{4cm}|}{\centering Subject and time dependence} \\
\cline{1-1} \cline{2-2} \cline{3-3} \cline{4-4}
\textbf{MMTB} & \textbf{0 (0)} & \centering \textbf{0 (0)} &  \textbf{0.03 (0.06)}\\
\hline
MMB & 0.09 (0.03) & 0.15 (0.01) & 0.04 (0.03)\\
\hline
mean-MMB & 0.22 (0.12) & 0.44 (0) & 0.44 (0)\\
\hline
median-MMB & 0.18 (0.13) & 0.44 (0) & 0.44 (0)\\
\hline
3d plaid & 0.44 (0.08) & 0.42 (0.1) & 0.39 (0.13)\\
\hline
\end{tabular}
\end{table}

\begin{table}
\caption{\label{tbl:simu_states_results}Recovery of states sequences, mean parameters and changepoints. Displayed are averages (standard deviation) over 30 datasets in simulation scenarios with time and/or subject dependence in state sequences. 
Preferred are: lower values of binder loss (BL) and mean absolute error (MAE); f-measure values closer to 1.}
\centering
\resizebox{\linewidth}{!}{
\begin{tabular}[t]{|l| c c c|c c c|c c c|>{}c|}
\hline
\multicolumn{1}{|c|}{ } & \multicolumn{3}{c|}{Time dependence} & \multicolumn{3}{c|}{Subject dependence} & \multicolumn{3}{c|}{Subject and time dependence} \\
\cline{2-4} \cline{5-7} \cline{8-10}
 & \CellWithForceBreak{measurement \\ partitions \\ (BL)}  & \CellWithForceBreak{state \\ parameters \\ (MAE)}  & \CellWithForceBreak{changepoints \\ (f-measure)} &  \CellWithForceBreak{measurement \\ partitions \\ (BL)}  & \CellWithForceBreak{state \\ parameters \\ (MAE)}  & \CellWithForceBreak{changepoints \\ (f-measure)} &  \CellWithForceBreak{measurement \\ partitions \\ (BL)}  & \CellWithForceBreak{state \\ parameters \\ (MAE)}  & \CellWithForceBreak{changepoints \\ (f-measure)}\\
\hline
MMTB & \textbf{0.08 (0.02)} & \textbf{0.73 (0.11)} & \textbf{0.73 (0.03)} & \textbf{0.09 (0.02)} & 1 (0.08) & \textbf{0.89 (0.02)} & \textbf{0.04 (0.02)} & \textbf{0.48 (0.11)} & \textbf{0.88 (0.04)}\\
\hline
SSTC & 0.09 (0.02) & 0.79 (0.07) & 0.63 (0.03) & 0.21 (0.1) & 1.89 (0.52) & 0.84 (0.04) & 0.1 (0.02) & 1.01 (0.09) & 0.74 (0.04)\\
\hline
tRPM & 0.12 (0.03) & 2.77 (0.34) & 0.22 (0.02) & 0.37 (0.05) & 3.31 (0.2) & 0.59 (0.02) & 0.2 (0.04) & 2.99 (0.35) & 0.34 (0.03)\\
\hline
MMB & 0.12 (0.02) & 1.23 (0.06) & - & 0.12 (0.01) & \textbf{0.88 (0.03)} & - & 0.09 (0.03) & 0.69 (0.06) & -\\
\hline
3d plaid & 0.31 (0.06) & 3.52 (1.07) & - & 0.6 (0.05) & 4.07 (0.6) & - & 0.45 (0.06) & 3.23 (0.85) & -\\
\hline
\end{tabular}}
\end{table}

\hypertarget{sec:discussion}{%
\section{Discussion}\label{sec:discussion}}

Motivated by the goal of learning multiple neurological profiles exhibited by subjects in electrophysiology and neuroimaging studies, we have proposed a Bayesian framework for temporal biclustering. The proposed methodology defines time-varying partitions of measurements nested within time-invariant partitions of subjects, thereby discovering a set of dynamic patterns common to multiple subjects.
We built on the framework of separate exchangeability, originally defined by \cite{kallenberg_representation_1989} and recently discussed by \cite{lin_separate_2021}. We adapted ideas from the temporal random partition model of \cite{page_dependent_2022} to the multi-subject setting.
We additionally developed and illustrated a blocked-sampling update of measurement partitions to efficiently explore the space of nested partitions and perform posterior inference via MCMC.
In our motivating application to fMRI data, changepoints in measurement partitions identified by our model largely coincided with the alternation of resting and activation blocks, and three groups of subjects were estimated with varying degrees of BOLD signal strength.
Our results on EEG data with 121 subjects further demonstrated the ability of our model to handle heterogeneity by identifying dynamic patterns shared by many subjects. Our experiments indicated good and consistent performance across simulated scenarios as well as improved estimation over existing methodologies.

Future work could consider extending the proposed model to allow for incorporating time-invariant covariates, such as subjects' characteristics, in the estimation of subject and measurement clusters. 
Moreover, in this paper we focused on the case where measurements are observed for all subjects at the same time intervals.
While this kind of data arises in many experimental settings, 
our framework could be adapted to model more general cases of longitudinal multivariate data. We envision that such a development could prove insightful in a number of real-world applications, e.g. in clinical studies monitoring disease progression among multiple patients.
The assumption that subjects with the same profile must share the same temporal partition for all measurements allows our model to obtain easy-to-interpret results, but it implies that the analysis must be restricted to the measurements which are relevant to the identification of subject profiles.
Endowing the model with more flexibility would be necessary if one wished to utilize a large number of measurements that may or may not be meaningful for estimating clusters of subjects. 
Random effects could be introduced to allow for more variability between subjects and/or measurements assigned to the same cluster, a strategy previously employed by \cite{kim_hierarchical_2006}. One could also consider using ideas from \cite{lee_nonparametric_2013} for excluding sets of irrelevant measurements, or borrowing from the literature on variable selection \citep{tadesse_handbook_2021} 
to allow measurements to contribute differently to the identification of subjects' partition.

For cases where several subjects undergo multiple experiments, it would be interesting to extend our model to study whether groups of subjects sharing the same dynamic patterns tend to be consistent across experiments, akin to the multi-view clustering problem recently considered by \cite{franzolini_conditional_2023} and \cite{dombowsky_product_2023}. 
While assuming that clusters of subjects are
invariant over time is a plausible assumption for the kind of short, single-task experiments
that we considered in this paper, one may need to allow subject partitions to vary over
time when performing temporal biclustering over longer time periods, or for experiments with
multiple heterogeneous tasks.

\paragraph{Acknowledgements} We wish to thank Sana Hussain for fMRI data collection and pre-processing. 
	
\bibliographystyle{plainnat} 
\bibliography{references-zotero.bib}       

\appendix

\newpage

\renewcommand{\thesection}{\Alph{section}}
\renewcommand{\thesubsection}{\Alph{section}.\arabic{subsection}}

\section{Supplementary material on posterior inference
}

We provide  details on all steps of the MCMC sampler that was outlined in Section 3.1 of the manuscript.


\subsection{Profile assignments} 
Conditional on the data $Y_i$ observed for subject $i$, on state-assignment sequences and on state-specific distributions, we can compute the likelihood of observations $Y_i$ under each possible profile assignment. 
Denoting the likelihood of subject $i$'s observations under profile $z$ by $\mathcal{L}_z(Y_i)$, we have

$$ \mathcal{L}_z(Y_i) = \prod_{r = 1}^{R}\prod_{t = 1}^{T} F \left( Y_{i, r, t} \mid \theta_{c^{(z)}_{r,t}} \right). $$

Then, the probability of assigning subject $i$ to profile $z$ can be derived by combining the likelihood of subject $i$'s observations under profile $z$ and the overall probability of profile $z$:

$$P\left(s_i = z \right) \propto \pi_z \mathcal{L}_z(Y_i).$$

The update of profile assignments has then a computational cost of $\mathcal{O}(R \cdot T \cdot Z)$, but this cost can substantially be reduced to $\mathcal{O}(Z)$  by computing $F \left( Y_{i, r, t} \mid \theta_{c^{(z)}_{r,t}} \right)$ in parallel across $R$ and $T$ .

\subsection{Profile probabilities} 
Conditional on each subject's profile assignment and on $\zeta$, the posterior distribution of profile probabilities is $\pi \sim \text{Dirichlet}\left( \dfrac{\zeta}{Z} + S_1, \dots, \dfrac{\zeta}{Z} + S_Z \right)$, where $S_z = \sum_{i = 1}^{N} \mathbbm{1}(s_i = z)$ is the number of subjects currently assigned to profile $z$. 

\subsection{State probabilities} 
Conditional on states assignments, we can resample overall state probabilities $\omega_0$ and profile-specific probabilities over states $\omega^{(z)}$ using auxiliary-variable methods developed for the HDP \citep{teh_hierarchical_2006}. 
Specifically, let $M^{(z)}_{k} =  \sum_{r = 1}^{R} \sum_{t = 1}^{T} \mathbbm{1}(c_{r,t}^{(z)} = k) \, \mathbbm{1}(\gamma_{r,t}^{(z)} = 0)$ be the number of times that state $k$ was sampled for each active profile $z$, that is a profile with at least one subject assigned (and set $M_{z, k} = 0$ for all $k$ whenever profile $z$ is inactive).
Given active profile $z$, each time state $k$ was sampled from $\omega^{(z)}$ it may have been sampled from the states available (i.e. already sampled) to profile $z$ or from $\omega_0$ (i.e. from an existing or a new table, in the framework of the CRP representation of the HDP).
To update $\omega_0$, we need to sample the auxiliary variables $T^{(z)}_{k}$, that is the number of times that state $k$ was sampled from $\omega_0$, across all profiles.
We set $T^{(z)}_{k} = 0$ whenever $M_{z, k} = 0$, otherwise we set $T^{(z)}_{k} = 1$ and, for $m = 2, \dots, M_{z,k}$ we do

$$
    T^{(z)}_{k} \leftarrow
    \begin{cases}
        T^{(z)}_{k} + 1 &\text{with probability } \dfrac{\alpha_{\omega} \omega_{0k}}{\alpha_{\omega} \omega_{0k} + m - 1} \\
        T^{(z)}_{k} &\text{o. w.}\\
    \end{cases}           
$$

Then, for each inner cluster $k \in \{1, \dots, K\}$, we set $$\bar{T}_{k} = \sum_{z = 1}^{Z} T_{z,k} $$ and we resample the expected frequencies of inner clusters from 
$\omega_0 \sim \text{Dirichlet} \left( \dfrac{\eta}{K} +  \bar{T}_{1}, \dots, \dfrac{\eta}{K} + \bar{T}_{K}\right)$.
For each outer cluster $z \in \{1, \dots, Z\}$, we can then resample its inner-cluster probabilities using $ \omega_z \sim \text{Dirichlet} \left( \phi \omega_{01} +  M_{z,1}, \dots, \phi \omega_{0K} + M_{z,K} \right)$.

\subsection{State persistence probabilities}
For each profile $z$ with at least one associated subject, for $t = 2, \dots, T$ let $G_{t}^{(z)} = \sum_{r = 1}^R \gamma_{r, t}^{(z)}$ be the number of measurements that persisted in the same state from time $t-1$ to time $t$.
By conjugacy, we can resample state persistence probabilities from a Beta posterior: $a_{t}^{(z)} \sim \text{Beta}\left(\alpha + G_{t}^{(z)}, \beta + R - G_{t}^{(z)} \right)$.

\subsection{Hyper-parameters of profile and subject probabilities}
To learn the posterior distribution of the variables controlling profiles and state probabilities (that is, respectively, $\varepsilon$ and $\upsilon$) we use a Metropolis Hastings (MH) step as in Algorithm 2, step (d-1) in \cite{fruhwirth-schnatter_here_2019}, sampling proposals from the prior distribution.
To encourage mixing, at each iteration we make a number of proposals equal to the expected number of proposals needed to achieve at least one acceptance.

\subsection{States-specific likelihood variables} 
The update of state-specific likelihood variables depends on the form of $F(\theta_k)$.
Conditional on subjects' profiles and on the state assignments of all measurements and time points for each profile, unless additional dependency is assumed by $F(\theta_k)$, the update will normally treat all observations assigned to state $k$ as exchangeable and will depend on a summary statistics of these observations.
In our simulations, we let $F(\theta_k)$ be a Normal distribution, place a conjugate Normal-Inverse-Gamma prior on $\theta_k = (\mu_k, \sigma^2_k)$ so that the conditional posterior can be sampled from directly.
In our real-data applications we let $F(\theta_k)$ be a more robust likelihood, the location-scale t-distribution $(\nu, \mu_k, \sigma_k)$ with fixed degrees of freedom $\nu = 3$ and with location $\mu_k$ and scale $\sigma_k$.
This is equivalent to assuming that $Y_{i,r,t} \mid s_i = z, c^{(z)}_{r,t} = k \ind \text{Normal} (\mu_k, V_{i, r, t})$ and $V_{i, r, t} \mid \sigma_k \iid \text{Inv-}\chi (\nu, \sigma^2)$, with the latter being the scaled inverse-chi-square distribution.
We sample from the posterior distribution of $\mu_k$ and $\sigma^2_k$ using the auxiliary-variable method detailed in \cite{gelman_bayesian_2013} and assuming conjugate priors $\mu_k \ind \text{Normal}(\mu_0, \sigma_0^2)$ and $\sigma_k^2 \sim \text{Gamma}(a_r, b_r)$.
For other choices of likelihoods, a Metropolis Hastings step may be needed when the conditional posterior cannot directly be sampled from.

\section{Supplementary material on application to neuroscience studies}
We show MCMC diagnostics of the temporal biclustering model applied to the experiments on neuroscience datasets in Section 4 of the manuscript.

\begin{figure}[h]
    \centering
    \includegraphics{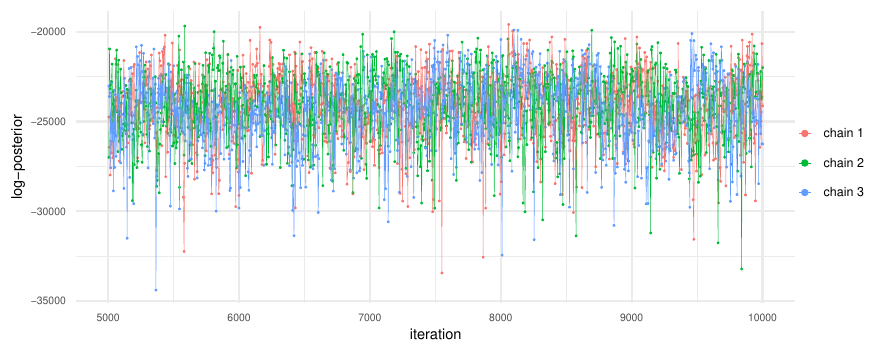}
    \caption{Joint log-posterior (up to a proportionality constant) of samples during the last 5000 iterations of our dynamic biclustering model on the fMRI dataset of Section 5.}
    \label{fig:fmri_diagnostics}
\end{figure}

\begin{figure}[h]
    \centering
    \includegraphics{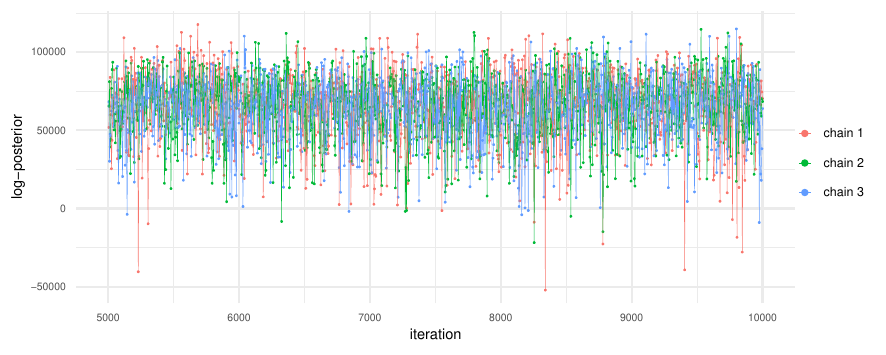}
    \caption{Joint log-posterior (up to proportionality constant) of samples during the last 5000 iterations of our dynamic biclustering model on the ERP dataset of Section 5.}
    \label{fig:eeg_diagnostics}
\end{figure}

The traceplots in Figure \ref{fig:fmri_diagnostics} indicate that the three MCMC chains that were ran on the fMRI datasets converged to an area of the posterior with similar joint probability. Note that a different random initialization was used for each chain. 
The same holds for the MCMC chains ran on the EEG dataset, as shown in Figure \ref{fig:eeg_diagnostics}.

\section{Supplementary material on simulation}
We detail the choice of hyperparameters and other model-fitting settings used in our experiments on simulated data in Section 5 of the manuscript.

When fitting the MMTB, SMTC and MMB models, in the notation defined in Section 2 and summarized in Figure 2 in the manuscript, we set $K = 20, d_1 = 50, d_2 = 100, \phi = 0.5$ and we let $\mu_k \iid \text{Normal}(0, 5)$ and $\sigma^2_k \iid \text{Inv-Gamma} (30, 30)$.
Further, for the MMTB and MMB, we let
$Z = 6, b_1 = 50, b_2 = 100$. 
For the MMTB model we set $\alpha = 10, \beta = 2$, while for the SMTC, that is fitted separately to each subject, we find that $\alpha = 3$ and $\beta = 2$ encourage smoothness in state sequences without overweighing the data.
For the tRPM, we set $\alpha$ and $\beta$ to the values giving best results, among the default and the values used for MMTB and SMTC.
The tRPM does not require a maximum number of states (i.e. measurement clusters) and uses a CRP model with base measure 1 for states’ probabilities. In the 3d plaid model~\citep{mankad_biclustering_2014}, a bicluster corresponds to  a subset of subjects \textit{and} a subset of measurements. We set the maximum number of biclusters to the total number of possible subsets of subjects and measurements, that is $2^{N} \times 2^{R}$. The 3d plaid model was also fitted with fewer biclusters, with similar results. For our models and the tRPM, we ran three MCMC chains for 20,000 iterations on each simulated dataset, discarding the first half as burn-in; we then combine samples from all chains to form approximate posterior distributions. 
Other settings of the tRPM and 3d plaid model were set to their default values. A single estimate for each parameter of the 3d plaid model was obtained by running its optimization algorithm.

\end{document}